\documentclass[twocolumn]{bmcart}

\usepackage{amsthm,amsmath,amssymb}
\RequirePackage{hyperref}
\usepackage[utf8]{inputenc} 
\usepackage{graphicx}

\startlocaldefs
\endlocaldefs

\begin{document}

\begin{frontmatter}

\begin{fmbox}


\title{Environmental unpredictability and inbreeding depression select for mixed dispersal syndromes}


\author[
   addressref={aff1,aff2},                   
   email={jhidalgo@onsager.ugr.es}   
]{\inits{JH}\fnm{Jorge} \snm{Hidalgo}}
\author[
   addressref={aff3,aff4,aff5},
   corref={aff3},
   email={rubiodecasas@ugr.es}
]{\inits{RRC}\fnm{Rafael} \snm{Rubio de Casas}}
\author[
   addressref={aff1},                   
   email={mamunoz@onsager.ugr.es}   
]{\inits{MAM}\fnm{Miguel \'A.} \snm{Mu\~noz}}


\address[id=aff1]{
  \orgname{Instituto Carlos I de F\'isica Te\'orica y Computacional and Departamento Electromagnetismo y F\'isica de la Materia, Universidad de Granada} 
  \postcode{18071}                                
  \city{Granada},                              
  \cny{Spain}                                    
}
\address[id=aff2]{%
  \orgname{Dipartimento di Fisica 'G. Galilei' and CNISM, INFN, Universit\'a di Padova},
  \street{Via Marzolo, 8},
  \postcode{35131}
  \city{Padova},
  \cny{Italy}
}
\address[id=aff3]{%
  \orgname{Departamento de Ecolog\'ia, Facultad de Ciencias, Universidad de Granada},
  \postcode{18071}
  \city{Granada},
  \cny{Spain}
}
\address[id=aff4]{%
  \orgname{Estaci\'on Experimental de Zonas \'Aridas, EEZA-CSIC},
  \street{Carretera de Sacramento s/n, La Ca\~nada de San Urbano},
  \postcode{04120}
  \city{Almer\'ia},
  \cny{Spain}
}
\address[id=aff5]{%
  \orgname{UMR 5175 Centre Ecologie Fonctionnelle et Evolutive, CEFE-CNRS},
  \street{1919 Route de Mende},
  \postcode{34293}
  \city{Montpellier cedex 05},
  \cny{France}
}

\begin{artnotes}
\end{artnotes}



\begin{abstractbox}

\begin{abstract} 
%
  \parttitle{Background} Mixed dispersal syndromes have historically
  been regarded as bet-hedging mechanisms that enhance survival in
  unpredictable environments, ensuring that some propagules stay in
  the maternal environment while others can potentially colonize new
  sites. However, this entails paying the costs of both dispersal and
  non-dispersal. Propagules that disperse are likely to encounter
  unfavorable conditions for establishment, while non-dispersing
  propagules might form populations of close relatives burdened with
  inbreeding.  Here, we investigate the conditions under which mixed
  dispersal syndromes emerge and are evolutionarily stable, taking
  into account the risks of both environmental unpredictability and
  inbreeding.
  \parttitle{Results} Using mathematical and computational modeling
  we show that high dispersal propensity is favored whenever temporal
  environmental unpredictability is low and inbreeding depression
  high, whereas mixed dispersal syndromes are adaptive under
  conditions of high environmental unpredictability, but more
  particularly if also inbreeding depression is small. Although pure
  dispersers can be selected for under some circumstances, mixed
  dispersal provides the optimal strategy under most parameterizations
  of our models, indicating that this strategy is likely to be favored
  under a wide variety of conditions. Furthermore, populations
  exhibiting any single phenotype go inevitably extinct when
  environmental and genetic costs are high, whilst mixed strategies
  can maintain viable populations even under such conditions.
  \parttitle{Conclusions} Our models support the hypothesis that the
  interplay between inbreeding depression and environmental
  unpredictability shapes dispersal syndromes, often resulting in
  mixed strategies. Moreover, mixed dispersal seems to facilitate
  persistence whenever conditions are critical or nearly critical for
  survival.
\end{abstract}




\end{abstractbox}
\end{fmbox}

\end{frontmatter}




\section*{Background} 

Organisms exist in ever-changing environments and need to surmount the
challenges posed by external variability.  When environmental
conditions vary unpredictably, the appropriate measure of evolutionary
success is not the average fitness across generations but its
geometric mean \cite{Dempster1955}. This is because population growth
is an inherently multiplicative process that is very sensitive to
occasional extreme values \cite{gardiner}. Thus, if organisms cannot
accurately predict or detect the most likely environment their
offspring will experience, they should hedge their bets by producing a
range of progeny phenotypes \cite{Starrfelt-Kokko2012}. For instance,
it has been argued that seed dormancy provides an instance of
parentally induced bet-hedging. Although it might entail germination
outside of the optimal time window, variable seed dormancy also
increases the probability for some individuals of the progeny to grow
under future favorable conditions \cite{simons-johnston,childs}. As a
consequence, probabilistic germination strategies --constituting a
sort of temporal bet-hedging-- are common in plants from semiarid
regions with unpredictable rainfall patterns \cite{satake}.

Similarly, spatial seed dispersal might enable plants to distribute
their progeny in different environments, minimizing the probability
that all of the seeds will disperse to unfavorable sites. Dispersal
appears to be particularly beneficial whenever conditions fluctuate in
time \cite{Hastings1983,Kisdi2002}. However, it entails a high risk,
as dispersers might end up in unfavorable patches. Consequently, it
has been posited that mixed dispersal strategies might emerge to
accommodate the risks of dispersal while still ensuring some of its
benefits. Several authors have supported this view and argued that the
production of progeny with contrasting dispersal abilities by a single
maternal genotype constitutes an instance of bet-hedging in
heterogeneous environments
\cite{hamilton-may,comins-hamilton-may,venable1985,zera-denno1997,Kisdi2002}

In addition to enabling sampling of new environments, dispersal
maximizes the probability of individuals encountering mating partners
of diverse genetic backgrounds. Locally dispersing individuals are
more likely to mate with relatives and therefore to produce inbred
progeny \cite{roze,guillaume-perrin2006}. Therefore, dispersal away
from the maternal site is expected to minimize inbreeding depression
and kin competition as well as to buffer variation in environmental
quality for the progeny \cite{bengtsson1978,hamilton-may}. In fact, it
has been argued that the prevention of inbreeding and kin competition
might be especially relevant for the evolution of dispersal strategies
\cite{Levin2003,roze,guillaume-perrin2006,Helene2000}.  Conversely,
high costs of dispersal, spatial unpredictability in environmental
conditions, and local adaptation are expected to select for limited
dispersal. Since kin competition and inbreeding avoidance are not
independent \cite{Levin2003,ronce2007,Helene2000}, the interaction of
these various forces can be summarized as a dynamical balance between
the avoidance of inbreeding and the costs and risks associated with
dispersal and environmental variability. However, to the best of our
knowledge, no study has previously modeled simultaneously the effect
of environment variability and inbreeding depression on the evolution
of dispersal.

\begin{figure}[tb]
\centering\includegraphics[width=0.95\columnwidth]{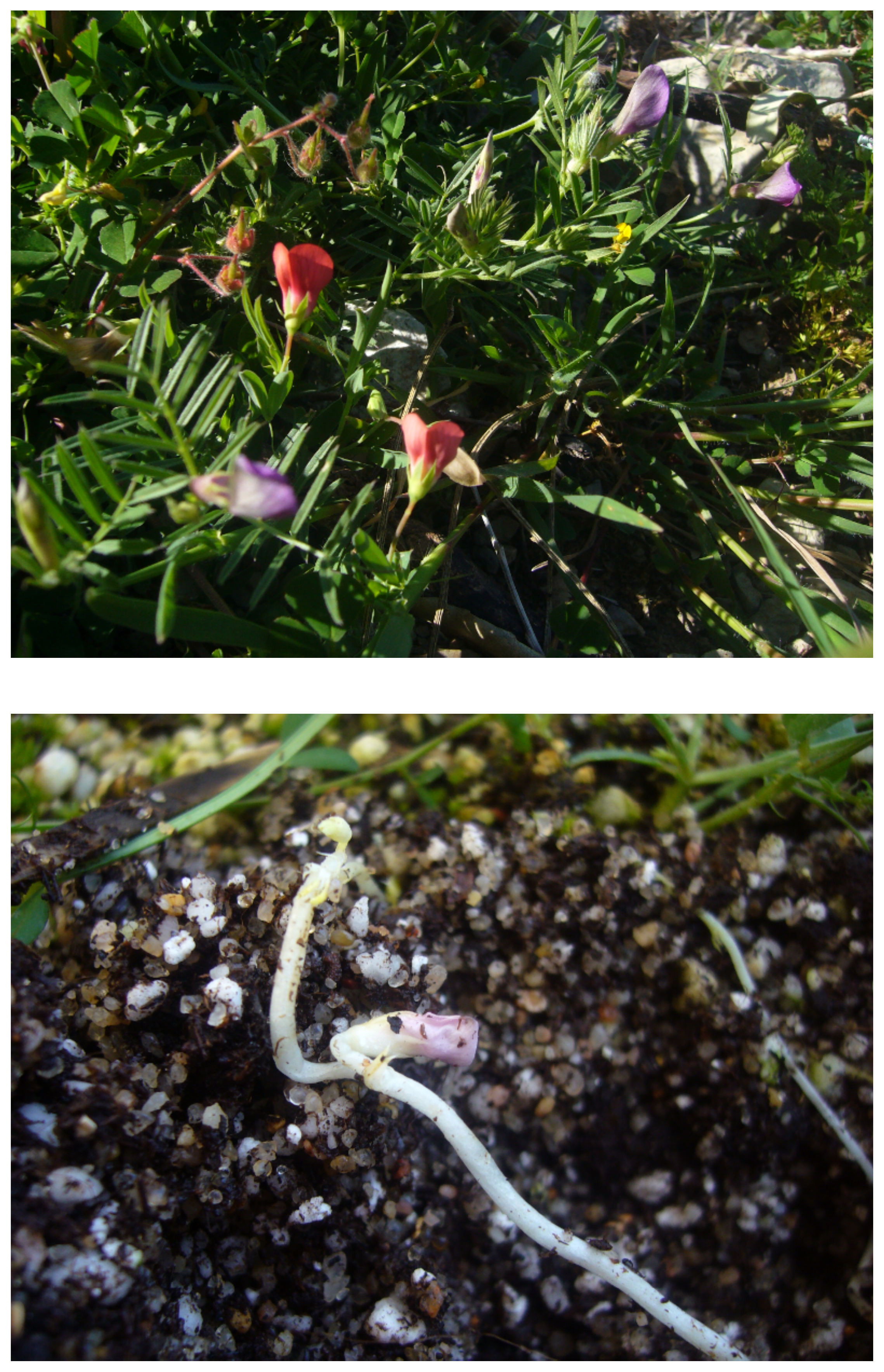}	
 \caption{\csentence{Two examples of amphicarpic plants exhibiting mixed dispersal syndromes}. 
 {(\bf Top)} Aerial open-pollinated flowers of {\it Lathyrus amphicarpos} (red) and {\it Vicia amphicarpa} (violet). {\bf (Bottom)} Subterranean self-pollinated flower of {\it Lathyrus amphicarpos}. Photos by Rafael Rubio de Casas.
 }
\label{fig:plants}
\end{figure}

A paradigmatic example of mixed dispersal syndromes is that of
heterocarpic plants that produce different types of seeds that vary in
their intrinsic dispersal propensity
\cite{venable1985,cheplick1987,mandak1997,imbert2002}.  In many cases,
the different dispersal phenotypes tend to be produced by flowers that
differ in their mating. For example, in many taxa, dispersing seeds
are produced by open-pollinated flowers while non-dispersing seeds are
produced by selfing flowers. This pattern can be observed in
amphicarpic plants that produce aerial chasmogamous (i.e.,
open-pollinated) and subterranean cleistogamous (i.e., strictly
self-pollinated) flowers such as {\emph{ Amphicarpaea bracteata,
    Amphicarpum purshii, Cardamine chenopodifolia, Lathyrus
    amphicarpos, Vicia amphicarpa, etc.}} as well as in other plants
that produce both chasmogamous and cleistogamous flowers, such as
\emph{ Agrostis hiemalis, Danthonia spicata, Impatiens capensis} and
\emph{Triplasis purpurea}
\cite{Schoen_Lloyd1984,imbert2002,barker2005,Herrera2009} (see
Fig.~\ref{fig:plants}). Although amphicarpy might seem a natural
history oddity, this association between open-pollinated flowers and
dispersing seeds and selfing flowers and non-dispersing seeds is an
extreme example of the positive evolutionary correlation between high
dispersal propensity and outcrossing and/or between selfing and
limited dispersal, predicted by several authors
\cite{guillaume-perrin2006,roze,Cheptou2009,Massol2011,Cheptou2012}.

Assuming evolution is likely to favor this correlation between
dispersal and the mating system, in the present paper we aim to
understand the evolution of dispersal strategies taking into account
{\it i)} the effect of temporal variability in environmental
conditions and {\it ii)} the effect of inbreeding depression. Our
model differs from previous studies of the correlation between
dispersal and the mating system --most notably from that of
\cite{Schoen_Lloyd1984}-- in that it explicitly incorporates both
environmental and genetic costs. Because both the risk of
establishment and inbreeding are expected to shape dispersal
strategies, we deem it necessary to develop a theoretical framework
for the evolution of dispersal that accommodates the two
processes.

Here, we develop mathematical and computational individual-based
models in which organisms live in a (discretized) two-dimensional
space where each individual is subjected to the demographic processes
of reproduction and death, with rates that depend upon temporarily
variable environmental conditions. Our models are inspired by
heterocarpic annual plants, i.e. by organisms with non-overlapping
generations with juvenile dispersal (\textit{sensu}
\cite{Massol-Debarre}) that can produce two different types of
propagules differing discretely in their dispersal propensity. As a
first approach, we consider a simpler, reductionist case that can be
assimilated to amphicarpy: plants produce two types of seeds that
differ both in their propensity for dispersal (dispersing versus
non-dispersing) and in their level of inbreeding (outcrossing versus
selfing). In this case, we posit a perfect correlation between
dispersal and the mating system that results in non-dispersing seeds
having a higher level of inbreeding than dispersing ones. Then, we
study a more general case in which the correlation between dispersal
and mating is not perfect. Plants still produce two types of
propagules, but inbreeding depression is derived as a function of the
proximity to relatives. In this case, we assume that mating occurs
only between individuals that are geographically close and inbreeding
depression ensues from the mating between individuals that are
genetically related.

In our modeling, dispersal phenotypes are assumed to be polymorphic,
with three possible dispersal syndromes: dispersing, non-dispersing,
and mixed, these last consisting in a combination of the other
two. This polymorphism varies among individuals and is expressed as an
intrinsic individual tendency to generate seeds/offspring of two
different types: dispersing or non-dispersing, which is coded in a
dispersal-propensity parameter $\alpha$. Our model allows us to
compute analytically and/or computationally --for any given set of
environmental conditions and inbreeding parameters-- the optimal value
of $\alpha$, which determines the level of dispersal propensity or bet
hedging. Additionally, we also develop an evolutionary (genetic)
algorithm that enables us to determine exactly what is the
evolutionarily stable dispersal syndrome under diverse circumstances.

The results of these models show that environmental variability is a
strong selective agent for dispersal and can facilitate the emergence
of purely dispersing or mixed dispersal syndromes. The specific
strategy that is selected for depends on the interplay between
inbreeding depression and environmental variability, although mixed
dispersal seems to be more favorable and robust under most
circumstances.

\section*{Methods}
\label{sec:model}

\begin{figure*}[tb]
\centering\includegraphics[width=0.95\textwidth]{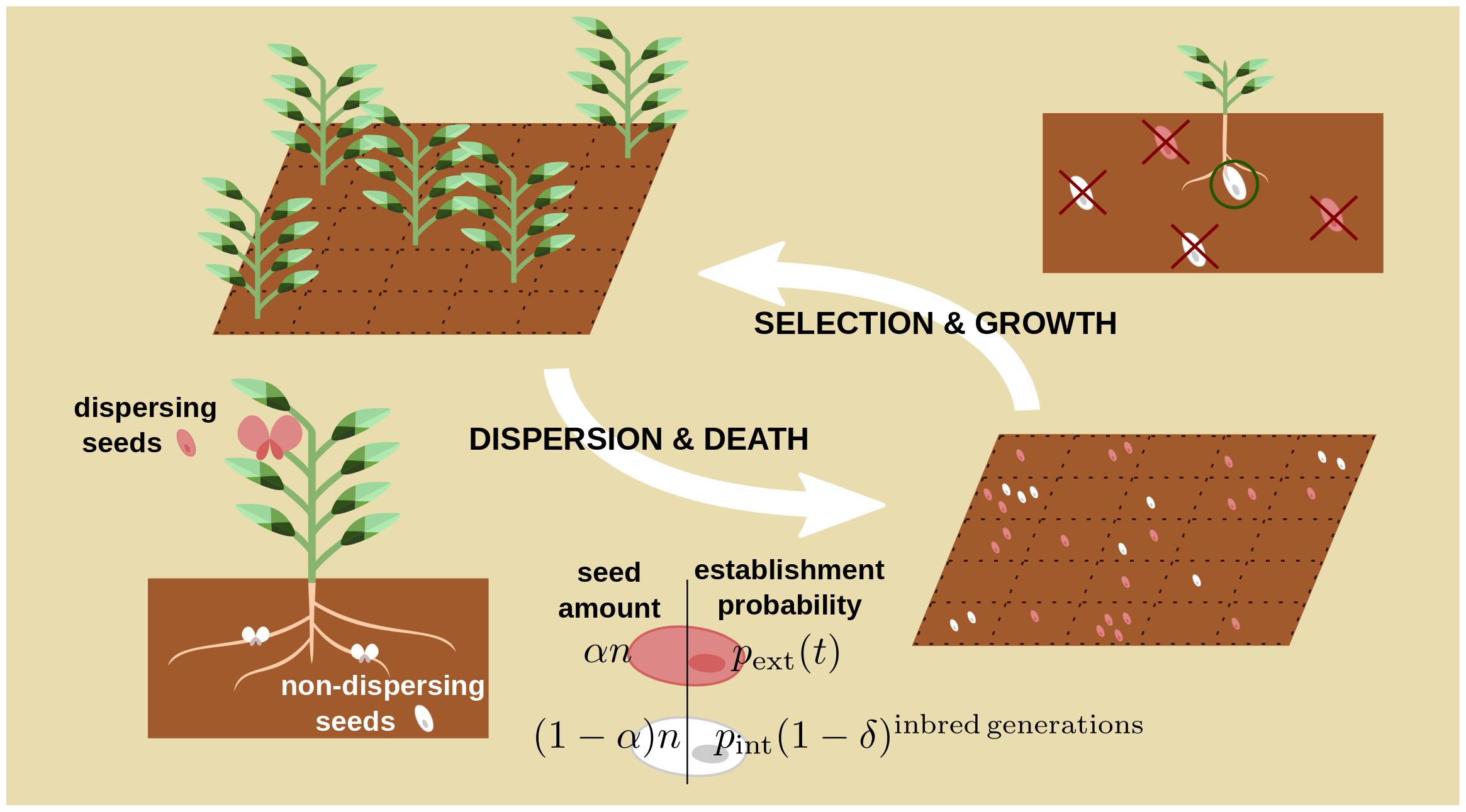}
 \caption{\csentence{Community of plants with mixed dispersal phenotypes}.
 Each plant is located at a cell of a square lattice of size $L\times L$.  Each individual plant produces the same fixed number of seeds, $n$; seeds can be of two different types: ``dispersing'' and ``non-dispersing'', marked in red and white respectively in the sketch.  Each produced seed is external/dispersing with probability $\alpha$, or internal/non-dispersing with probability, $1-\alpha$, where $\alpha$ is the dispersal propensity parameter. After reproduction, all adult plants are removed from the community. Regardless of their origin, dispersing seeds can randomly arrive to any cell in the lattice where they get established with probability $p_\mathrm{ext}(t)$ which is environment-dependent. On the other hand, non-dispersing seeds can only establish themselves at the maternal location or in its adjacent lattice cells. Then, for cells with more than one established seed, one of these is chosen at random and the rest die. The establishment probability $p_\mathrm{int}$ of non-dispersing seeds is assumed to be independent from environmental variability (and thus, it does not depends on time). In the simplest case, dispersing seeds are produced by outcrossing, whereas non-dispersing seeds are the product of selfing. Thus, their quality, $q$, is reduced after each inbreeding event by a penalization factor $q\rightarrow(1-\delta)q$. In our formulation, inbreeding depression is approximated in a manner that can be assimilated to the interaction among many slightly deleterious alleles that affect the trait independently (e.g., \cite{Charlesworth}). The number of alleles determining inbreeding is assumed to be $n\rightarrow\infty$. Therefore, it is always proportional to homocigosity and accumulates multiplicatively with inbreeding events. Outcrossing events are expected to eliminate homozigosity and to reset $\delta=0$ and $q=1$, so dispersing seeds are always assumed to not have any inbreeding depression.
 We also study a more general scenario in which selfing is not restricted to a specific dispersal syndrome, but in which inbreeding affects the quality of seeds produced by individuals that mate with relatives (see Appendix A). In this case, inbreeding is proportional to the kinship between mating partners. This generalized model is less restrictive in that there is no assumption of a perfect correlation between dispersal and mating, but leads to qualitatively similar conclusions.
 }
\label{fig:model}
\end{figure*}

We present a simple and parsimonious individual-based model in which a population of plants develops in time through the processes of birth, reproduction, competition, and death. Each individual/plant lives at a fixed site on a two-dimensional square lattice of size $L\times L$. Periodic boundary conditions are assumed to minimize contour effects. In order to account for local competition for resources and space, occupancy is restricted to a maximum of one plant per site (see Fig. \ref{fig:model}).

At each discrete time-step $t$ --which represents a reproductive
cycle, i.e.  one year in the case of annual plants-- all occupied
lattice sites are emptied, i.e. generations are assumed to be
non-overlapping, and new plants emerge from existing seeds following
some dynamical rules that we specify in what follows.  Each individual
plant produces the same, fixed, number of seeds, $n$, but these can be
of two different types/morphs: ``dispersing'' and ``non-dispersing'',
respectively. Dispersing seeds travel to distant sites --for
simplicity, we assume that they can end up randomly in any location
within the lattice-- whereas non-dispersing seeds can only stay in the
maternal or adjacent sites. The relative fraction of these two types
is modulated by the so-called dispersal propensity parameter $\alpha$:
with probability $\alpha$, each of the produced seeds is dispersing,
or, complementary with probability $1-\alpha$,
non-dispersing. Initially, we take $\alpha$ as a constant, while
allowing for variability in other model parameters (e.g., the degree
of inbreeding depression and environmental variability). In a second
step, we study the case in which $\alpha$ is dynamically self-tuned in
the community.

Even if, for the sake of simplicity, we assumed that the reproductive
output --i.e. the total number of seeds produced by each individual
plant-- is constant, fitness differs among maternal plants because the
probability of establishment are constrained by environmental
conditions and inbreeding depression and thus the actual contribution
to the next generation is individual-dependent.

Each morph follows a different type of dynamics: 
\begin{itemize}

\item Dispersing seeds establish at a randomly selected site with a
  probability $p_{\mathrm{ext}} (t) \in [0,1]$ --which is a
  fluctuating random variable assumed to depend upon environmental
  conditions and to be equal at each time step for all sites in the
  lattice-- or, alternatively, they can be lost with complementary
  probability $1-p_{\mathrm{ext}} (t)$.  For simplicity, we take
  $p_{\mathrm{ext}}(t)$ to be an uncorrelated random variable
  --freshly extracted at each discrete time step-- with uniform
  distribution in $[\bar p_{\mathrm{ext}}-\sigma, \bar
  p_{\mathrm{ext}}+\sigma]$ with the constraint that $\sigma <
  \min(\bar p_{\mathrm{ext}}, 1-\bar p_{\mathrm{ext}})$.
\item Non-dispersing seeds are assumed not to be influenced by
  environmental variability, but to suffer from inbreeding
  depression. In particular, individual plants are equipped with an
  individual {\it quality} parameter $q$, which is inherited by the
  seeds they produce. The actual establishment probability of
  non-dispersing seeds is $q \times p_{\mathrm{int}}$, where
  $p_{\mathrm{int}}$ is the maximal value it can take.  The quality
  $q$ --modulating the establishment probability-- is reduced by a
  factor $(1-\delta)<1$ every time there is selfing (i.e., a
  reproductive event resulting in a non-dispersing seed), or instead,
  it is reset to $q\rightarrow 1$ by outcrossing (i.e., when a
  dispersing seed is produced). We set $q(t=0)=1$ for all individuals
  in the community. Future, more complex approaches should include
  also environmental dependency in the probability of establishment of
  non-dispersing seeds.
\end{itemize}
After all reproductive events and establishments at each timestep (year), only one seed is randomly selected for reproduction at each cell (if the number of seeds in that cell is $\geq 1$) while the rest are removed from the community.

It is important to keep in mind that the current model is a parsimonious one, restricted to the simplest syndrome, i.e., one in which the association between dispersal and mating is immediate and perfect and in which temporal environmental variability affects only dispersing seeds. 
To test the robustness of our results in the absence of of a perfect correlation between dispersal and mating, in Appendix A we also study a more general scenario in which the probability of selfing is independent of the dispersal syndrome, but in which inbreeding depression affects seeds produced by individuals that mate with relatives proportionally to the kinship between mates. 
Still, non-dispersing seeds produce plants that are more likely to coexist with their relatives, and are indirectly more affected by inbreeding than those individuals coming from dispersed seeds. Consequently, the generalized model leads to results that are similar to those of the simpler version. 

Given that all our computer simulations are run considering finite and closed populations, extinction is always possible --even in cases with a relatively high stationary density-- as a consequence of demographic fluctuations. Once all individuals have disappeared from the community, the system reaches a stationary state and remains quiescent indefinitely. Note that, strictly speaking, other stationary states of the system are quasi-stationary; i.e. they reach a steady state conditioned on the system not having reached the quiescent state in which the population goes extinct \cite{gardiner}. Extinctions can then be classified in two different categories: 
\begin{itemize} \item \emph{Deterministic} extinctions, which occur with certainty after a given characteristic time, which grows slower than linearly with system size; these occur when the system lies at the ``absorbing'' phase.
 \item \emph{Accidental} extinctions, which correspond to catastrophic demographic fluctuations, the probability of which rapidly decreases --exponentially-- with system size. These only occur --as a result of finite-sizes-- when the system is in its ``active'' phase.
\end{itemize}
In order to determine --for a given set of parameter values-- in which phase the system lies, we measured computationally the mean-extinction time $T$ as a function of the linear system size $L$. $T$ grows exponentially or algebraically with $L$ in the active phase, while it converges asymptotically to a constant value in the absorbing one.  Additionally, to measure the quasi-stationary density in an efficient way, we re-activated any iteration reaching the quiescent state (i.e., extinction)  by setting it to a small but non zero density (i.e., re-introducing by hand a few individual plants in the community). Albeit admittedly ad-hoc, this computational trick allows us to avoid limiting our statistics to active realizations of the model, and leads to similar results than other more sophisticated exact methods \cite{Dickman-quasistationary}.

Our working hypothesis is that, in unpredictable environments or when
inbreeding depression is significant, mixed dispersal strategies ($0 <
\alpha <1$) might lead to higher individual fitness than either of the
single phenotype syndromes.  To test this, we first develop a
preliminary study of the stationary density $\rho$ for different
values of the dispersal parameter $\alpha$, while keeping all other
parameters fixed. We find that for each choice of parameters, there
exists a specific value $\alpha^*$ for which the population density is
maximized. However, selection for dispersal is frequency-dependent,
which makes it impossible to approximate the ESS using optimality
criteria \cite{Kisdi1998}. Consequently, we implemented an
evolutionary approach in which $\alpha$ is not a constant, kept fixed
across the whole population but an inheritable variable.  In
particular, we defined a genetic algorithm similar to that of
\cite{goldberg,gros} in which each individual has its own dispersal
syndrome, as encoded in a specific value of its parameter $\alpha$;
this value is transmitted to its progeny with a stochastic (Gaussian
distributed) variation of zero mean and $\nu$ standard deviation. In
biological terms, $\nu$ can be understood as the rate of mutation,
recombination and other sources of variation in heritable traits
across generations (referred to as mutation rate hereafter). In this
evolutionary version of the model, individuals with low fitness tend
to become extinct, while the space they leave empty becomes
progressively occupied by fitter individuals (individuals with a
higher probability of establishment).  The outcome is a population
with some averaged (quasi)steady-state density, $\rho$, and a
well-defined averaged value of $\alpha$ across the community, that we
call $\bar{\alpha}$, and some variance around these mean values.  This
modeling exercise does not aim at representing a realistic
evolutionary process. It is just an effective dynamic in which
individuals in the population self-optimize their dispersal strategies
across generations as a result of competition and mutation.

\section*{Results}
\label{sec:results}
\subsection*{Single phenotype cases: $\alpha=0,1$}
\label{subsec:sim_phases_pure}

As a first step, we studied the behavior of the population when it exhibited any of the two single phenotype syndromes, i.e. the value of the dispersal propensity parameter was fixed to either $\alpha=1$ or to $\alpha=0$, respectively.  To proceed, we computed through numerical simulations (starting from a fully occupied lattice) the stationary population density, $\rho$, as a function of parameter values: $\bar p_{\mathrm{ext}}$ and $\sigma$ for dispersing seeds ($\alpha=1$) and the couple $ p_{\mathrm{int}}$ and $\delta$ for the non-dispersing case ($\alpha=0$).   In most of what followed, we fixed the original establishment probabilities for both types of seeds to be identical, $p_{\mathrm{int}}=\bar p_{\mathrm{ext}}$, but our main results do not change, at least qualitatively, for asymmetric values of the establishment probabilities.  

\begin{figure}[tb]
\centering\includegraphics[width=0.95\columnwidth]{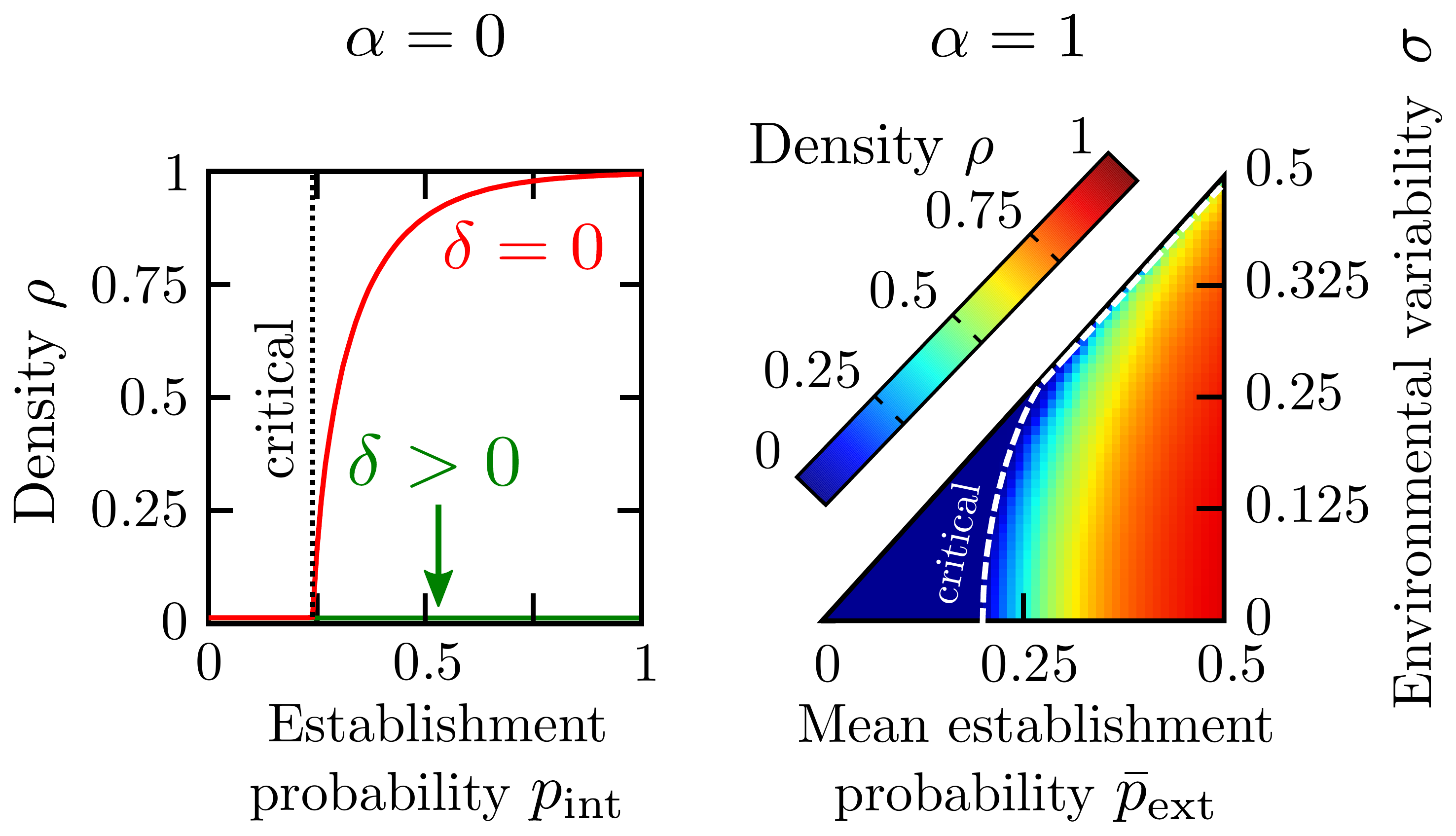}
 \caption{ \csentence{Stationary density of single phenotype syndromes}.
 Both single-phenotype dispersal syndromes, $\alpha=0$ (left) and $\alpha=1$ (right), exhibit a phase transition between an \emph{absorbing} phase where the population always becomes extinct, and an \emph{active} phase, with a sustained stationary density. 
 For the non-dispersing syndrome (left) the population always becomes extinct with non-null inbreeding depression ($\delta>0$, green curve), while there is a phase transition at $p_{\mathrm{int}}\simeq0.24$ if $\delta=0$ (red curve).
 In the dispersing syndrome (right), the critical point increases as a function of the degree of environmental variability, $\sigma>0$, i.e.  variability is detrimental to population density, expanding the absorbing phase (dark blue region) at the expense of the active one. The critical line has been computed with the analytical approach described in Appendix B for an infinite system, $L=\infty$.  To compute the stationary densities, we iterated for $10^4$ generations, and averaged over the last $10^4/2$ steps; averages over at least $100$ independent realizations were performed. Parameters are set to $L=100$ and $n=5$.
} 
\label{fig:phases_pure}
\end{figure}

The left panel of Fig. \ref{fig:phases_pure} shows the stationary
density for the single non-dispersal syndrome ($\alpha=0$). In the
absence of inbreeding depression ($\delta=0$), there is a critical
point located at $p_\mathrm{int}^c\simeq0.24$ 
However, for any non-zero inbreeding depression ($\delta>0$), the purely non-dispersive phenotype $\alpha=0$ is doomed to extinction regardless of the establishment probability $p_\mathrm{int}$.  

Fig. \ref{fig:phases_pure} (right panel) illustrates the (quasi)stationary density for the single dispersal syndrome ($\alpha=1)$ for different choices of the environmental parameters $\bar p_\mathrm{ext}$ and $\sigma$: a continuous phase transition separates an absorbing phase (region above the dashed line) in which all plants become (deterministically) extinct and an active one, in which a non-trivial (quasi)steady state is reached (region below the dashed curve). In the absence of environmental variability (i.e. $\sigma=0$) the critical point at which the phase transition occurs lies at a \emph{critical} establishment probability $\bar p_\mathrm{ext}^c(\sigma=0)=1/n=0.20$ (i.e. when persistence is ensured by the production of one viable offspring by each maternal individual): large probabilities entail non-trivial steady states, and small ones lead ineluctably to extinction. Similarly, as $\sigma$ increases larger establishment probabilities are needed to sustain a non-trivial density. 

Measurements of averaged extinction times were then used to confirm the location of the critical lines in both cases. Furthermore, the critical lines can be analytically calculated as shown in detail in Appendix B. Although the two single-phenotype cases we have just discussed are overly simplistic and results are somewhat predictable, they provide a useful reference to frame the dynamics of mixed strategies.

\subsection*{Mixed dispersal syndromes}
\label{subsec:sim_model1}

As a second step, we explored how the quasi-stationary density,
$\rho$, changes as a function of the dispersal propensity parameter
$\alpha$ for various environmental and inbreeding conditions (i.e. for
diverse choices of $\sigma$ and $\delta$.

\begin{figure}[tb]
\centering\includegraphics[width=0.95\columnwidth]{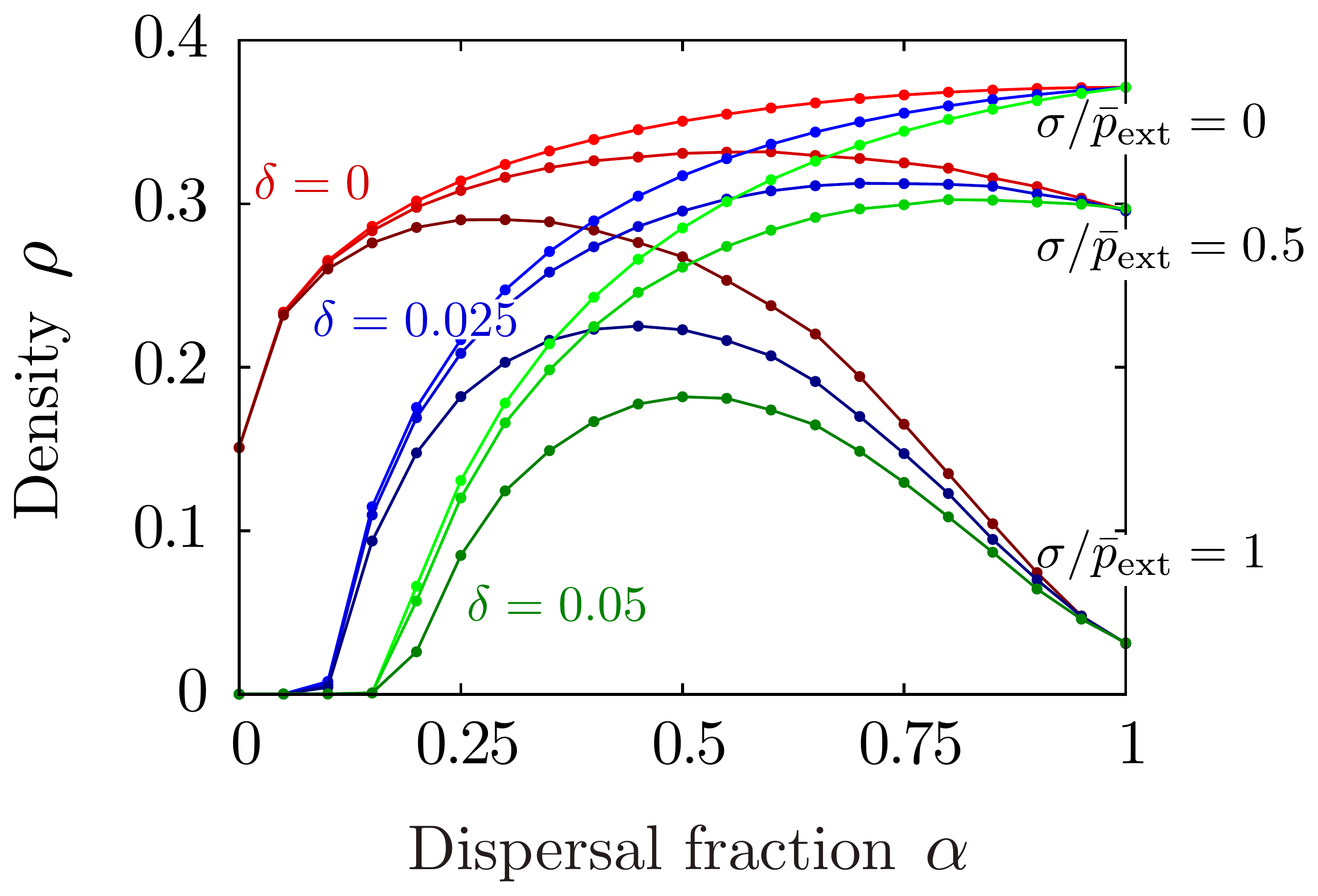} 
\caption{
\csentence{Stationary density for the mixed dispersal syndromes as a function of the dispersal propensity parameter $\boldsymbol{\alpha}$}. 
Parameters of the single phenotype syndromes are set to $p_{\mathrm{int}}=\bar p_{\mathrm{ext}}=0.25$ and $n=5$. Linear system size $L=100$ (total size $N=10000$).  Stationary density for three different values of inbreeding depression (represented by red, green and blue curves, respectively) and environmental variability (darker shades indicate higher environmental unpredictability; labeled at $\alpha=1$).  Both $\delta$ and $\sigma$ tend to reduce the population density in the pure strategies ($\alpha=0$ or $\alpha=1$, respectively) but, remarkably, relatively large densities can be attained by populations with mixed syndromes even in the presence of inbreeding depression and environmental unpredictability. Although the specific values of $\delta$ and $\sigma$ are not intended to be biologically realistic, change in these parameters illustrates qualitatively the consequences of different genetic and environmental costs. Note that points at parameter values $\alpha=1$ and $\sigma=0.25$ correspond to the absorbing region (see the calculation $\bar p_\mathrm{ext}^c$ in Appendix B for $\sigma=p_\mathrm{ext}^c$) however measurements of the (quasi)stationary density give small positive values, which decrease to zero for larger system sizes (not shown).
}
\label{fig:rho-alpha}
\end{figure}

Results are summarized in Fig. \ref{fig:rho-alpha} (again, simulations
were run fixing $\bar p_\mathrm{ext}=p_\mathrm{int}$).  First, it can
be noticed that, in the absence of environmental variability (i.e. for
$\sigma=0$), the stationary density grows monotonically with $\alpha$,
indicating that the purely dispersing syndrome leads to higher
populations densities. In our models, the system is not saturated and
therefore dispersal tends to be favored because it enables the
colonization of new sites. As the values of $\sigma$ increase, the
density attained by populations exhibiting the purely dispersing
syndrome decreases whereas populations with mixed syndromes
(intermediate values of $\alpha$) attain larger stationary densities
than the pure dispersal strategy. The relative advantage of mixed
syndromes is even more conspicuous when inbreeding depression is also
significant ($\delta>0$). For instance, for $\delta=0.025$ and
$\sigma=0.15$, the curve is non-monotonous and has a parabolic-like
shape, with local minimal densities at $\alpha=0$ and $\alpha=1$ and
with a maximum at $\alpha^*\approx 0.6$.

Moreover, we observed a broad region in parameter space ($\delta>0,\sigma\gtrsim0.2$) where the population became extinct when individuals spread through either of the single phenotype syndromes but survived when the dispersal syndrome was mixed.

\subsubsection*{Extinction times}
\label{sec:ext-times}
\begin{figure}[hb!]
\centering
\includegraphics[width=0.95\columnwidth]{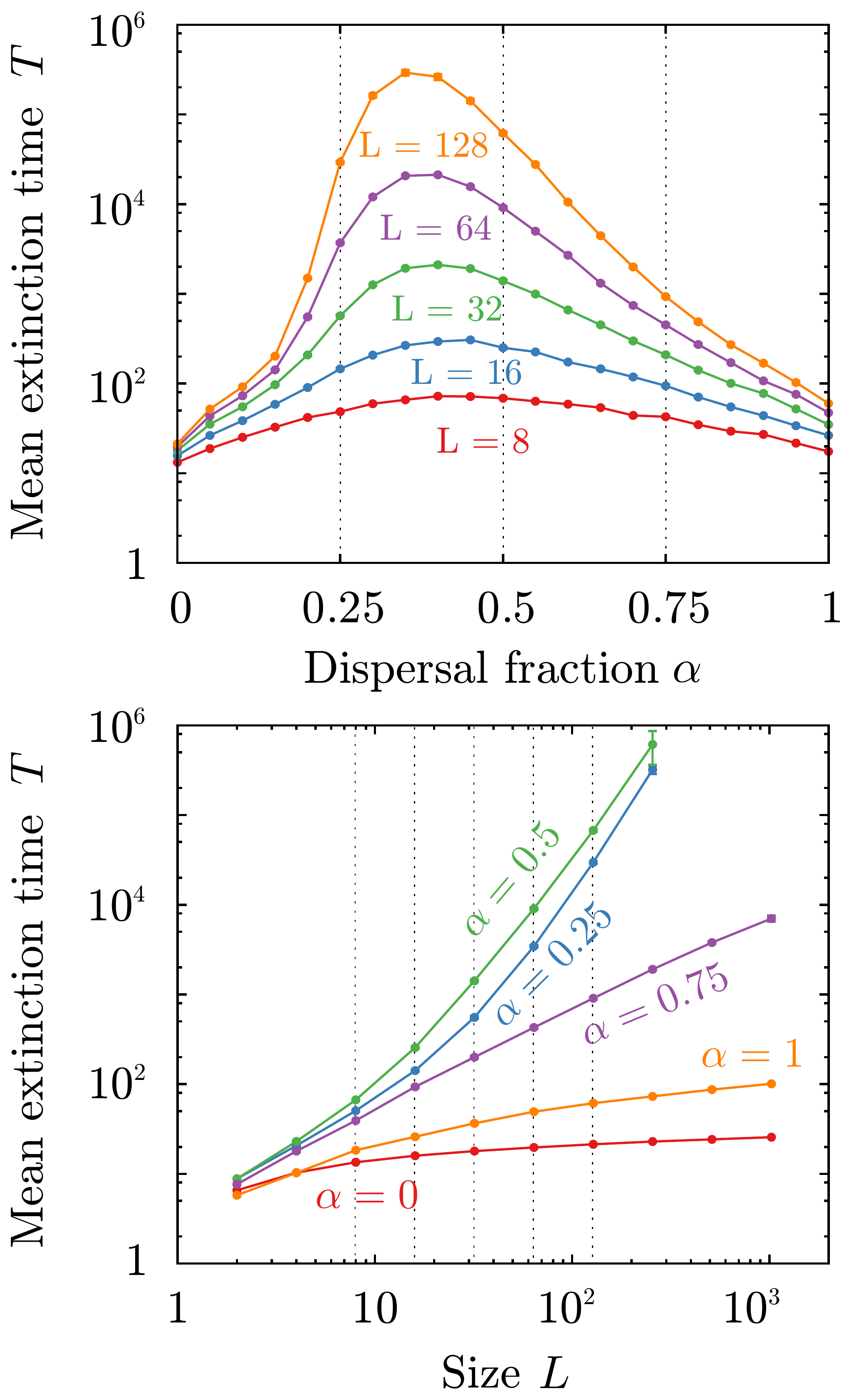}
\caption{
\csentence{Mean extinction time, $T$, for mixed dispersal syndromes as a function of the dispersal fraction $\alpha$ (top) and of linear system size $L$ (bottom)}.
Parameters of the single phenotype syndromes are set to $p_{\mathrm{int}}=\bar p_{\mathrm{ext}}=0.25$, $n=5$, $\delta=0.05$ and $\sigma=0.25$, corresponding to the absorbing phase (see Fig. \ref{fig:phases_pure}). Each point was computed averaging over $10^3$ realizations of the simulations. Most error bars are smaller than point size. Dashed lines have been included to facilitate comparisons across panels. (Upper panel) Extinction time is always maximum when $0.25<\alpha<0.5$ and converges to $\alpha^*\approx 0.4$ for large sizes of the system).  (Lower panel) Populations exhibiting any of the single-phenotype strategies ($\alpha=0$ and $\alpha=1$) are in the absorbing phase, as indicated by the downward curves on the log-log scale, i.e., their extinction is unavoidable (deterministic extinction).  Conversely, intermediate values of $\alpha$ lead to extinction times that grow exponentially with system size, and thus maintain populations in the active phase.}  \label{fig:ext-times} \end{figure}

To confirm the positive effect of mixed dispersal strategies, we
performed the following computational experiment: we tuned our free
parameters $\sigma=0.25$ and $\delta=0.05$ to make both dispersing and
non-dispersing syndromes nonviable on the long term, leading
ineluctably to extinction (i.e. the system was forced towards the
absorbing phase). Starting from a fully occupied system, $\rho=1$, we
computed the mean extinction time for different values of $\alpha$ and
various linear system sizes $L$. Results are shown in
Fig. \ref{fig:ext-times}.  Fig. \ref{fig:ext-times} (upper panel)
shows the averaged extinction time $T$ as a function of $\alpha$ for
system sizes $L=8,16,32,64$ and $128$. In all cases, larger systems
have longer extinction times. However, although $T$ barely varies for
single phenotype syndromes ($\alpha=0,1$), extinction times rapidly
increase for mixed dispersal syndromes (e.g. $\alpha=0.5$).
Fig. \ref{fig:ext-times} (lower panel) illustrates the extinction time
as a function of the system size for values of the dispersal parameter
$\alpha=0, 0.25, 0.5, 0.75$ and $1$. While single dispersal syndromes
($\alpha=0,1$) exhibit a slower-than-polynomial dependency (concave
function on a log-log scale), extinction times increase exponentially
for mixed dispersal syndromes (convex function on a log-log scale).
These evidences support that, even in cases where both pure syndromes
result in deterministic extinction in relatively short times, mixed
ones can be active and lead to stable populations, surviving for
exponentially large times.

We used a mathematical analysis to understand how mixed dispersal syndromes are able to enable higher population densities and much longer extinction times. These calculations are described in detail in Appendix B. In a nutshell, we computed --in the simplest possible scenario-- the averaged growth rate, $G$, as a function of $\alpha$ for small population densities. In this case, the (non-linear) effects of competition and saturation can be safely neglected, rendering the calculation amenable to exact analytical solutions. The sign of $G$ determines whether the population tends to shrink and disappear ($G<0$), or, instead, to grow and survive ($G>0$). Our results show that $G(\alpha)$ is a non-linear function such that, even for parameter values for which both single dispersal syndromes would lead to extinction ($G(\alpha=0,1)<0$), some mixed syndromes can result into a positive growth rate ($G(\alpha)>0$), allowing for long-term persistence. 

\subsubsection*{Optimal strategy}
\label{subsec:sim_model2}
In the evolutionary version of the model, the dispersal propensity parameter $\alpha$ dynamically evolves though the processes of competition, inheritance and mutation. Here, each individual plant has its own value of $\alpha$, which is transmitted to its progeny (i.e. to each seed, then to a new plant) with a small Gaussian mutation with zero mean and standard deviation $\nu$, imposing that $\alpha=0$ (resp. $\alpha=1$) whenever $\alpha<0$ ($\alpha>1$).

Initially, plant dispersal syndromes are drawn from some arbitrary probability distribution $P(\alpha,t=0)$ (all plants are initially considered to have $\alpha=1/2$). At each generation, $P(\alpha,t)$ is dynamically modified and, eventually, achieves a stationary shape $P(\alpha,t\rightarrow\infty)$, identified with the evolutionary stable strategy (ESS). At this point, we compute the (quasi)-stationary density and the mean-value of $\alpha$ in the community, that we call $\bar{\alpha}$ . Runs that led to  accidental extinction were re-activated with 
a value of $\alpha$ chosen at random from those of the generation immediately preceding extinction.  After verifying that stationary values of density and $\alpha$ were largely independent of $\nu$ (if $\nu$ is sufficiently small) we fixed $\nu=10^{-3}$.

\begin{figure}[htb!]
\centering\includegraphics[width=0.95\columnwidth]{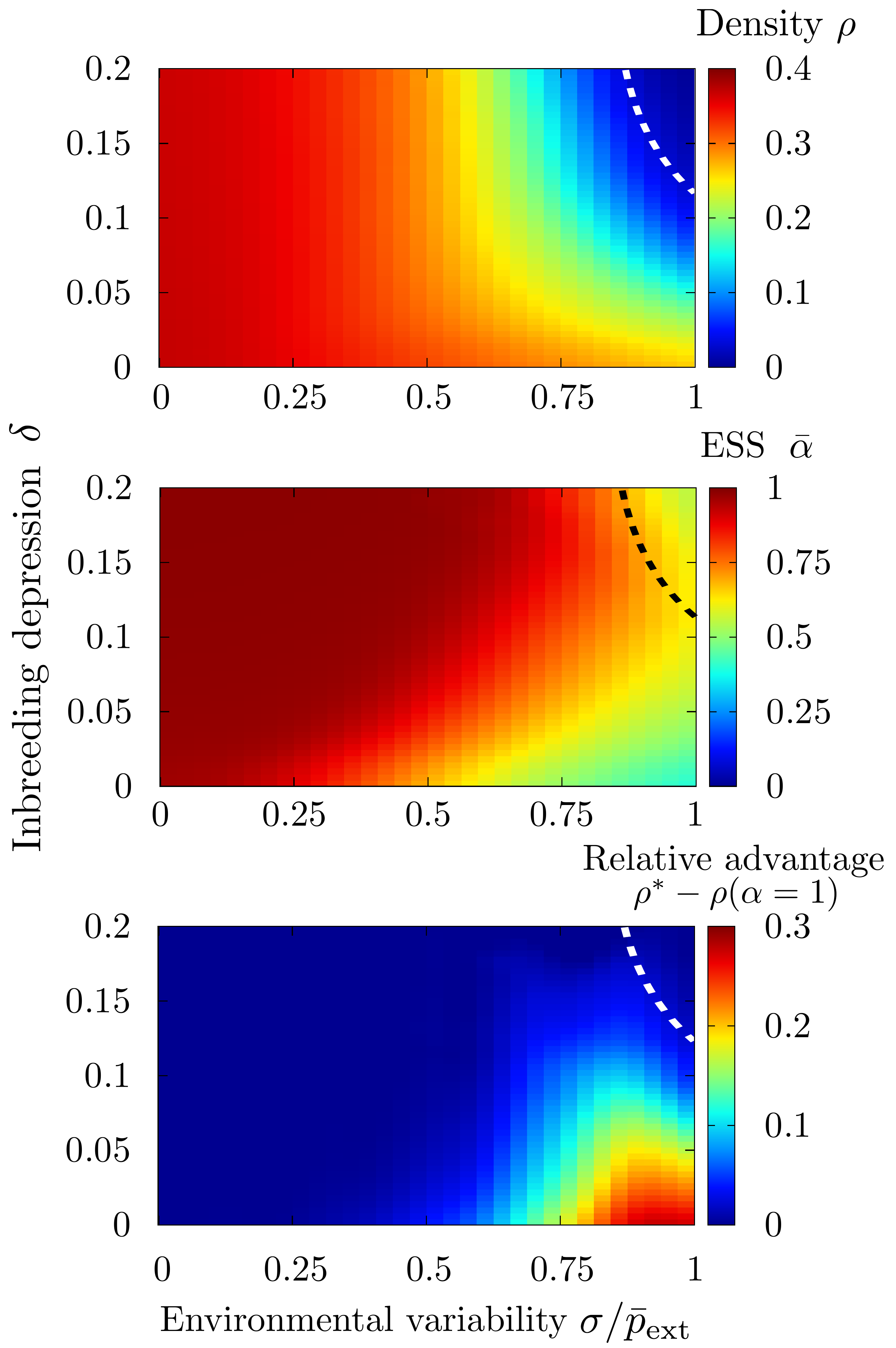} 
 \caption{\csentence{Optimal strategy}.
 The mean dispersal propensity parameter $\alpha$ is self-tuned dynamically in a community of individuals through evolutionary dynamics, based on the genetic algorithm of \cite{goldberg,gros}. (Upper panel) Color plot of the stationary density as a function of the inbreeding depression $\delta$ and the environmental variability $\sigma$; the dashed line represents the critical line separating absorbing and active phases in the quasi-stationary ensemble (i.e., the system is not allowed to go fully extinct) and is established as the parameter combinations that ensure a stationary density $\ = 1\%$. Although individual plants have the possibility of developing mixed strategies, the population always becomes extinct whenever both $\delta$ and $\sigma$ are high; i.e. under very adverse conditions. (Middle panel) Average value of $\alpha$ across the population in its steady state. The dispersing syndrome is favored whenever the environmental variability is low  $\sigma<0.05$ and/or inbreeding depression is high (red region). The ESS corresponds to mixed dispersal when environmental variability is significant (orange to green region). The parameter space in which the ESS corresponds to predominantly non-dispersing syndromes is very narrow (hardly visible dark blue region in the lower right hand corner) and requires inbreeding depression to be negligible and environmental variability to be large. 
 (Lower panel) Relative advantage of the mixed dispersal syndromes: stationary density of the optimal mixed dispersal syndrome, $\rho^*$ (obtained from the genetic algorithm), minus the density for the fixed dispersal syndrome.
 Parameters have been set to $L=100$, $p_{\mathrm{int}}=\bar p_{\mathrm{ext}}=0.25$, $n=5$ and $\nu=10^{-3}$, and averages are performed over the last $10^5/2$ steps in 10 independent simulations iterated for $10^5$ generations.
  }
 \label{fig:GA} 
 \end{figure}

Results for the mean density and mean value of $\alpha$ are shown in Fig. \ref{fig:GA} (upper and middle panel, respectively). The active phase (i.e. to the left of the dashed line representing the critical transition) can only be attained under certain combinations of environmental variability and inbreeding depression; the ESS always leads to non-saturated population ($\rho < 0.4$).  All ESS have values of $\bar{\alpha}$ in the interval $[\approx 0.35, 1]$ in the active phase.  These solutions have relatively small standard deviations  (below $0.04$) around their mean values, indicating that there is little heterogeneity in $\alpha$ across the population in the steady state.  
Remarkably, the values of $\bar{\alpha}$ are very close to the previously obtained values $\alpha^*$ for which the population density is maximized for all cases (i.e. for all values of $\delta$ and $\sigma$).
In summary, our evolutionary models resulted in populations that were either strictly dispersing or had mixed dispersal. Moreover, the ESS was always homogeneous across the population.
It is worth noting that optimal values of $\bar{\alpha}$ could be calculated even for populations in the absorbing phase. These values of $\bar{\alpha}$ describe the strategies opposing the strongest possible resistance to extinction; i.e. the dispersal strategies of the last survivors in a community proceeding towards deterministic extinction.

Finally, it is important to mention that the way in which we have implemented an evolutionary dynamics into our model is just a simple way to define and measure the ESS, rather than a realistic evolutionary model.

As a final step, we tried to quantify the relative evolutionary advantage provided by mixed dispersal syndromes with respect to pure phenotypes. This was estimated as the difference between the steady state density of a population exhibiting the optimal dispersal strategy (ESS) and the maximum density attainable by a  population exhibiting a pure dispersing syndrome: $\rho^*-\rho(\alpha=1)$. No comparison with the single non-dispersal phenotype was computed because this strategy always leads to extinction, $\rho(\alpha=0)=0$, for any non-zero value of inbreeding depression. Results for different environmental and inbreeding parameters values, $\sigma$ and $\delta$, are reported in the lower panel of Fig. \ref{fig:GA}. The increase in population density provided by the emerging optimal mixed syndrome is much higher in the lower right part of the Figure, i.e. for relatively small values of inbreeding depression ($\delta<0.05$) and relatively large values of the environmental variability $\sigma>0.2$.

As explained in Appendix B, the gain derived from mixed dispersal strategies can be expressed mathematically: the exponential growth rate is much larger for intermediate values of $\alpha$ than for extreme values of the dispersal propensity, close either to $0$ or to $1$.  The advantage provided by mixed dispersal is visible even under the simplifying assumptions used in the mathematical calculations.

\section*{Discussion}
\label{sec:discussion}

The results derived from our models showed that dispersal syndromes
can be the direct outcome of the interplay between inbreeding
depression and environmental temporal variability. Depending on the
specific strength of these two forces, the optimal dispersal syndrome
can either ensure a high dispersal propensity, very limited dispersal,
or a mixed situation in which individuals employ a combination of both
of the previous strategies.

According to our analyses, pure populations of non-dispersers can only
be viable on large timescales (i.e., reach non-trivial steady-state
densities $\rho > 0$) in the complete absence of inbreeding depression
($\delta=0$). Similarly, environmental variability reduced the steady
state density of pure dispersing populations but, contrary to
non-dispersers, viable populations of dispersers can exist under a
wide range of environmental uncertainty. Under rather
generic conditions, populations with the pure dispersing syndrome
tended to perform better than pure non-dispersing populations. This
result was not unexpected, as in the absence of environmental
variability our model penalized non-dispersal through inbreeding
depression. However, it highlights the role of inbreeding depression
in the evolution of dispersal, in agreement with the results obtained
by other authors \cite{roze,guillaume-perrin2006,Cheptou2012}.

Generally speaking, optimal dispersal strategies appear to be represented by mixed syndrome in which plants produce simultaneously dispersing and non-dispersing seeds; populations with dispersal fraction $0.25<\alpha<0.75$ attained higher densities and were viable for longer whenever environmental conditions were highly fluctuating (i.e., for any $\sigma\geq 0.2$). Moreover, mixed dispersal appears to enable positive growth rate and long term survival even under conditions for which both single dispersal phenotypes would lead to extinction. This is congruent with the findings of Jansen and Yoshimura \cite{jansen99}, who showed that populations can persist in an environment consisting of sink habitats, if offspring are randomly distributed over the two of them. Our results support this idea and indicate that mixed syndromes can be advantageous due to the benefits of bet-hedging through multiple co-existing complementary strategies. 

In all of our computational analyses, patch size was finite, and thus populations could possibly go extinct --following stochastic demographic collapse-- relatively easily. More importantly, according to our models, population extinction was unavoidable, regardless of patch size, whenever inbreeding depression or extreme environmental variability occurred.  In particular, dispersing syndromes led to accelerated extinction if the population was subject to adverse environmental conditions for several generations. These situations might or might not take place in real-world scenarios, where meta-population dynamics can buffer the effect of local extirpations through migration from other sources \cite{Meta1,Meta2}. However, even disregarding immigration, populations exhibiting mixed strategies had significantly longer extinction times, and these grew very fast with patch size. This result is likely contingent on temporal autocorrelation of the environment, and if the latter is positive, extinction is expected to be faster under unfavorable conditions for any given phenotype. However, even if no phenotype can guarantee unlimited survival in finite patches, expected extinction times are drastically enhanced if organisms display mixed dispersal syndromes.
 
The dynamical/adaptive version of our model also supported the hypothesis that populations with mixed syndromes are more resilient. In this version of the model, dispersal propensity was not fixed but rather dynamically self-organized to its optimal value. Results showed that mixed syndromes appeared to provide the highest population densities and the longest population life-spans. In our formulation, non-dispersing had an intrinsic penalization for any non-null value of inbreeding depression. In spite of this relative advantage of the dispersing phenotype, pure dispersal was found to be the evolutionary stable strategy only if environmental unpredictability was low, especially if inbreeding depression was also low. Other authors have predicted that mixed dispersal is adaptive in heterogeneous environments \cite{venable1985,snyder2006,snyder2011}. Our models indicated that this is indeed the case, but that the optimal dispersal strategy is also contingent on inbreeding depression, and that if the latter is significant and dispersal and mating are closely correlated, pure dispersers might have an advantage over mixed dispersers even in heterogeneous environments.

Even though environmental and genetic costs influence the evolution of dispersal, environmental variation appears nevertheless to be specially relevant for the emergence of mixed dispersal strategies, particularly when the requirement of a perfect correlation between mating and dispersal is released. This result might be seen as contradicting the findings of the model put forward by \cite{Schoen_Lloyd1984}. These authors showed that mixed mating/mixed dispersal can become the ESS when each type of propagule provides a clearly different advantage; higher establishment in the case of non-dispersing, inbred seeds and lower sibling competition in the case of outbred, dispersing seeds, and these predictions are independent of environmental variability. In our models, sibling competition and other kin selection mechanisms are implicitly incorporated into the inbreeding depression term (i.e., they can be regarded as deleterious consequences of being in close proximity to kin), while environmental variability affects the probability of establishment of dispersing seeds alone. Thus, our model can be in fact regarded as a corroboration of Schoen and Lloyd's findings \cite{Schoen_Lloyd1984}; mixed syndromes are beneficial when there are simultaneous costs to dispersal and coexistence with kin. However, our predictions deviate from Schoen and Lloyd's in that we anticipate a relatively wide range of conditions under which mixed dispersal can be selected for, and predict that, in the (near) absence of deleterious interactions with kin, costs of dispersal might be enough to select for mixed dispersal.

According to our more reductionist model, mixed syndromes provided the
ESS under environmental and mating conditions of high environmental
unpredictability and/or low inbreeding depression. Moreover, under
those same conditions they had a significantly higher fitness than any
other alternative phenotype.  The results of the general model showed
an even wider region in which mixed syndromes were favored, mostly due
to the lesser influence of inbreeding depression.

The purely non-dispersing syndrome was almost never found to provide
an ESS.  Moreover, the robustness of the optimal phenotypes was very
high, as indicated by the low variation around the optimal ESS.  This
well-defined optimal ESS seemed to rule out even the emergence of
non-dispersing syndromes in populations with multiple phenotypes. In
the case of the more general model, relatively low values of $\alpha$
were observed under high environmental variation, but those values
were always $\geq0.25$. This is somewhat surprising, as plants with
monomorphic non-dispersal syndromes do exist. For instance, there are
several taxa that produce all their seeds underground (i.e.,
geocarpic) such as the peanuts (\emph{Arachis} spp.), \emph{Trifolium
  subterraneum} or \emph{Macrotyloma geocarpum} and non-dispersal has
been shown to be an ESS by different models
\cite{LudwigLevin1991,snyder2011}. According to these models,
non-dispersing phenotypes are adaptive whenever the probability of
establishment away from the maternal site is lower than within the
maternal site (i.e., whenever there is local adaptation).  We have not
explored in detail these scenarios in our models. However, naively,
one could anticipate that non-dispersal syndromes could emerge as ESS
by considering the extreme case in which $p_{\mathrm{int}} \gg \bar
p_{\mathrm{ext}}$ and the inbreeding depression is negligible, $\delta
\approx 0$ or if the models allowed for purging or included a
transmission advantage of selfing \cite{Husband_Schemske}. However,
although the adaptive value of limited dispersal needs further
investigation, this sort of phenotype is rare in nature and might be
seen as exceptional \cite{Levin1991}.

Finally, it is important to note that the way in which we implemented evolutionary dynamics into our calculations provide just a simple way to define and measure the ESS, rather than realistic evolutionary models. For instance, we assumed that both types of reproduction and dispersal have the same costs for the maternal plant and that the two types of propagules produce identical individuals in the absence of inbreeding depression or environmental variability. However, the empirical and theoretical literature have shown that non-dispersing and dispersing propagules require different resource allocation from the mother plant and can produce different progeny \cite{imbert2002}. Moreover, it is possible that mixed syndromes are not easy to develop. It could be expected that the mutations necessary to generate a mixed system, i.e.,mutations from $\alpha=0,1$ to $\alpha\neq 0,1$ would have high pleiotropic costs and be hard to gain. This sort of functional polymorphism entails the coexistence of two distinct phenotypes within a single individual, each phenotype comprising a suite of traits (e.g., flower and fruit tissues, architectural traits, etc.) with their own development and maintenance particularities. However, a biologically plausible model able to incorporate these additional complexities is beyond the scope of the present work.

\section*{Conclusions}
\label{sec:conclusion}
Our results show that although dispersal can be selected for under a
wide range of conditions, mixed dispersal syndromes should be favored
in systems that are exposed to highly unpredictable
environments. Moreover, mixed syndromes seem to ensure the viability
of populations for longer periods of time, particularly under limiting
conditions of environmental variability and inbreeding depression.

\section*{Appendix A. Generalization of the model: absence of an explicit correlation between dispersal and mating}
\label{sec:general_model}
The models presented so far are based on a perfect correlation between dispersal and the mating system, as only non-dispersing seeds experience inbreeding depression. In this section, we show that this condition can be relaxed, leading to qualitatively similar conclusions, i.e., that mixed dispersal strategies are selected for by environmental unpredictability and enable population persistence in the presence of inbreeding and dispersal costs.

In a more general scenario, we suppose that: (i) inbreeding modifies the quality of seeds, $q$, regardless of their dispersal strategy, which modulates their establishment probability $p_\mathrm{ext}(t)\rightarrow q\times p_\mathrm{ext}(t)$ and $p_\mathrm{int}\rightarrow q \times p_\mathrm{int}$, for dispersing and non-dispersing seeds respectively; (ii) we assume that mating is more likely to occur between individuals that are spatially close. Therefore, in the presence of inbreeding depression, the quality of the seeds produced by a plant is a function of its proximity to relatives and of its genetic similarity to these relatives, i.e., if a plant coexists with its close relatives, it is likely to mate with them and to produce seeds of bad quality. Assuming inbreeding depression is the product of multilocus interactions among many ($n\rightarrow\infty$)   moderately or slightly deleterious alleles and that the number of said alleles that are homozygous in an inbred genotype will determine its fitness \cite{Charlesworth}, the effect of inbreeding depression can be modeled as follows: given a plant $i$, we compute the quality of its seeds, $q(i)$, as:
\begin{equation}
 q(i) = 1- \left\langle e^{-{a(i,j)}/{\delta'}}\right\rangle_{j\in n.n.(i)}
 \label{eq:genetic}
\end{equation}
where $\langle\cdot\rangle_{j\in n.n.(i)}$ represents the average over the plant's immediate neighbors, $\delta'$ is the inbreeding depression parameter (see below), and $a(i,j)$ is the minimum number of past generations in which individuals $i$ and $j$ have a common ancestor, i.e., the degree of kinship between them, so that $a(i,j)=1$ if they come from the same mother, $a(i,j)=2$ for a common grandmother but different mothers, and so on. As eq. \ref{eq:genetic} is not defined in the case of a plant with zero neighbors, we take a maximum inbreeding depression that still can be modulated by $\delta'$, imposing for such case $q(i) = 1-e^{-{0.5}/{\delta'}}$. This means that in the absence of neighbors, selfing is complete and homozigosity maximal. However, if any neighbors are present, cross-pollination is assumed regardless of the kinship between individuals.

Eq. \ref{eq:genetic} affects both dispersing and non-dispersing seeds, and therefore there is no assumed correlation between mating and dispersal. However, non-dispersing seeds produce plants that are more likely to coexist with their relatives, and, indirectly, become more affected by inbreeding than those individuals coming from dispersed seeds.

To understand the role of the new parameter $\delta'$, we can analyze the extreme cases $\delta'\ll1$ and $\delta'\gg1$: for low values of $\delta'$, the exponential function in eq. \ref{eq:genetic} rapidly decreases to 0, and $q\rightarrow1$; this limit represents a situation in which individuals can coexist with close relatives without suffering inbreeding depression. On the other hand, high values of $\delta'$ represent the case in which individuals are subject to high inbreeding depression even when mating with distantly related partners, as $q\rightarrow0$. In conclusion, $\delta'$ plays a similar role to the parameter $\delta$ in the model introduced before.


\begin{figure}[htb]
\centering\includegraphics[width=0.95\columnwidth]{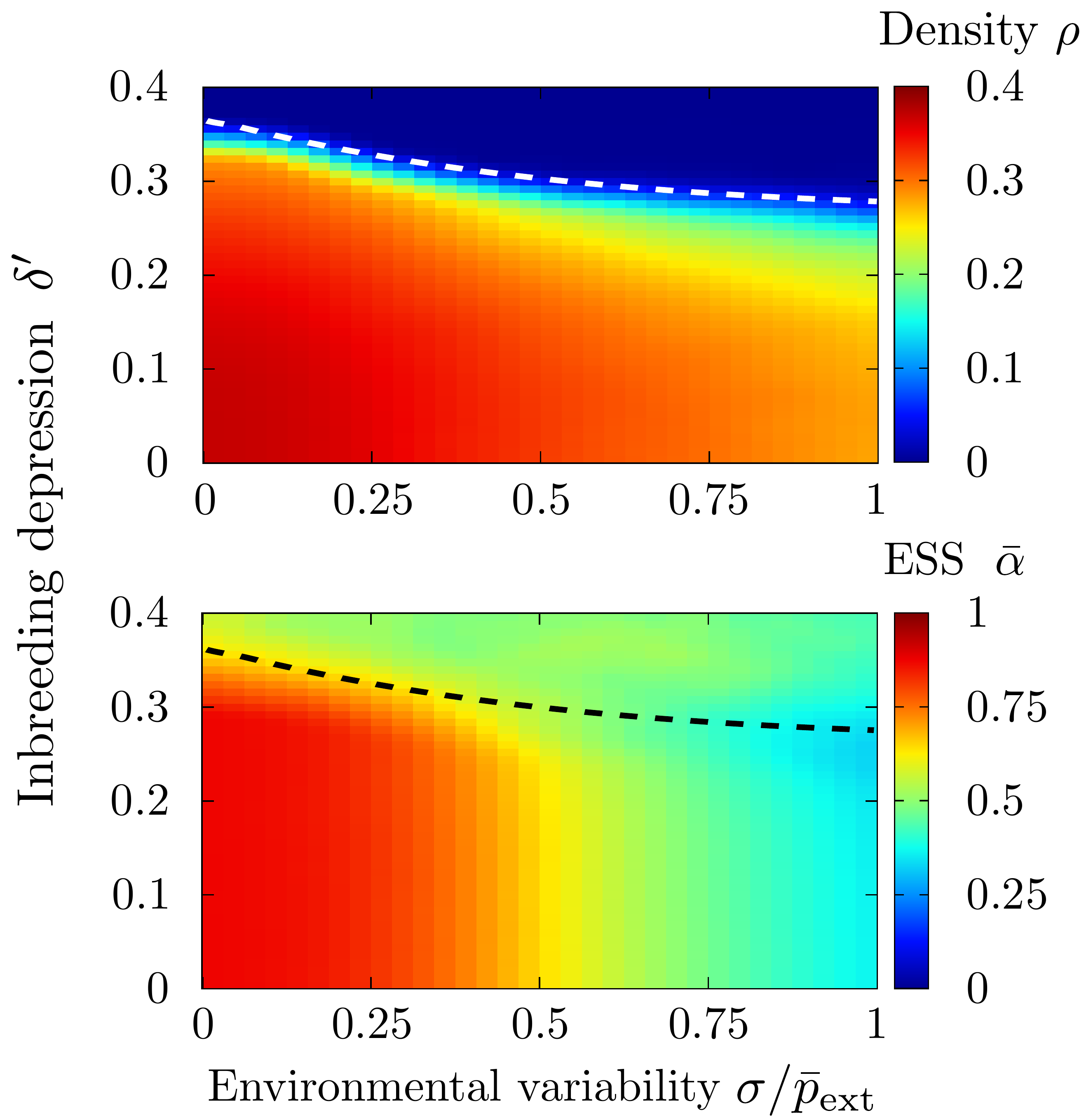}
\caption{\csentence{Evolutionary stable strategies in the generalized model}.
(Upper panel) Stationary density of the population as a function of the inbreeding depression $\delta'$ and the environmental variability $\sigma$; the dashed lines separate the absorbing from the active phase. (Lower panel) Average value of $\alpha$ across the population in its steady state. Parameters have been set to $L=100$, $n=5$, $p_{\mathrm{int}}=\bar p_{\mathrm{ext}}=0.25$ and $\nu=10^{-2}$, and averages are performed over the last $10^4/2$ steps in simulations iterated for $10^4$ generations.} 
\label{fig:GA_genetic}
\end{figure}

In order to calculate the optimal strategy, we also developed an evolutionary implementation of the general model. Fig. \ref{fig:GA_genetic} shows the results obtained by letting the mean dispersal propensity parameter $\alpha$ self-tune dynamically in a community of individuals through evolutionary dynamics based on the genetic algorithm of \cite{goldberg,gros}. This figure is equivalent to Fig. \ref{fig:GA} and shows the effect on population density of inbreeding depression and environmental variability (upper panel) and the corresponding dispersal ESS for each combination of parameters (lower panel). As observed in the simple model, mixed syndromes are selected for in the proximity of the critical point and the dispersing syndrome is favored whenever the environmental variability is low (red region). However, in this case the ESS corresponds to mixed dispersal whenever environmental variability is $\sigma >0.1$ (orange to light blue region) for any value of $\delta'$. Generally speaking, the ESS values of $\alpha$ are lower in this case than in the simpler model because the inbreeding depression affects both dispersal and non-dispersal propagules and thus the quality of non-dispersing individuals is not so severely penalized compared to dispersing ones. In other words, although under this generalization inbreeding affects all individuals, its influence on dispersal is less significant. As a result, $\alpha$ is largely independent from $\delta'$ for any $\delta'\leq0.25$. However, under high values of $\delta'$ we again observe that the populations surviving under stressful conditions exhibit mixed dispersal syndromes (low density populations near the critical line in Fig. \ref{fig:GA_genetic} ). In spite of the large parameter region in which mixed dispersal constitutes the ESS, purely non-dispersing syndromes ($\alpha\simeq0$) are still not selected for under any parameterization, as evidenced by the absence of dark blue regions in the lower panel of Fig. \ref{fig:GA_genetic} . 

It is worth noting that even though the generalized model presented here no longer assumes a perfect correlation between mating and dispersal, it still considers only the case of a discrete dispersal polymorphism (i.e. seeds are either dispersed or implanted locally). Additionally, only dispersed seeds are affected by environmental fluctuations. We leave for future research further generalizations of the model considering continuous rather than discrete dispersal kernels as well as more extensive parameterizations of environmental variability.

\section*{Appendix B. Analytical approach}
\label{sec:theory}
In addition to the computational results described in the main text, we developed a mathematical model in order to better understand the mechanisms by which mixed dispersal affected population density and extinction time. This approach is, in essence, an analytical evaluation of the population average exponential growth rate of the computational model. This growth rate can be defined as: 
\begin{equation} G = \lim_{t\rightarrow\infty} \frac{1}{t} \log \frac{N(t)}{N(0)}. \end{equation}
This quantity is negative for shrinking populations, positive for growing ones, or zero in the marginal or critical case; therefore, the sign of $G$ can be regarded as an indicative proxy of the fate of the population.  We calculated $G$ as a function of the environmental and inbreeding parameters for different types of homogeneous populations with a fixed value of $\alpha$: i) Single phenotype dispersing syndrome $(\alpha=1)$, ii) Single phenotype non-dispersing syndrome $(\alpha=0)$ and iii) mixed syndromes ($0<\alpha<1$). The mathematical analyses can be performed under the following approximations: 

\begin{itemize} 

\item Saturation effects --which become important at large densities preventing the population from growing infinitely-- are neglected. Therefore, the following calculations provide valuable information only for the dynamics of low-density populations, allowing in particular the determination of critical points, but not to study stationary states.

\item System sizes are sufficiently large, so that, statistical deviations from mean values can be safely neglected.

\item Spatial correlations are not included; i.e. all sites are assumed for simplicity to be nearest neighbors.

\end{itemize}
In the terminology of statistical mechanics these approximations, taken together, constitute a linearized ``mean field'' approach \cite{Marro}.

\subsection*{Dispersal syndrome ($\alpha=1$)}
In this approximation, $n p_\mathrm{ext}(t)$ is the average number of established seeds per plant at any given generation $t$.
Thus, starting from a population $N(t=0)$ individuals after $t$ generations the population size becomes
\begin{equation}
 N(t) = n p_\mathrm{ext}(t) N({t-1}) = ... = \left( \prod_{i=1}^t  n p_\mathrm{ext}(i) \right) N(0).
\end{equation}
Using this expression and the definition of $G$,
\begin{equation}
 G = \frac{1}{t}\sum_{i=1}^t \log\left(n p_\mathrm{ext}(i) \right) = \left\langle{\log\left(n p_\mathrm{ext}(t)\right)}\right\rangle_t,
\end{equation}
where $\langle \cdot \rangle_t$ represents the temporal average over generations using the probability distribution of $p_\mathrm{ext}(t)$.  Observe that, as $\frac{1}{t}\sum_{i=1}^t \log\left(n p_\mathrm{ext}(i) \right)= \log( \Pi_{i=1}^t \left(n p_\mathrm{ext}(i) \right)^{1/t})$, the overall growth rate coincides with the geometric mean of growth rates across generations; i.e. population growth rate is a multiplicative process \cite{Dempster1955}.

In the particular case in which $p_\mathrm{ext}$ is uniformly distributed in the range $[\bar p_\mathrm{ext}-\sigma, \bar p_{ext}+\sigma]$ (and assuming that all possible values of $p_\mathrm{ext}$ have been homogeneously sampled for sufficiently large times $t$), $G$ can be explicitly calculated as
 \begin{eqnarray}
G &=& \int_{\bar p_\mathrm{ext}-\sigma}^{\bar p_{ext}+\sigma}  dp_\mathrm{ext} \frac{1}{2\sigma} \log\left( n p_\mathrm{ext} \right) \nonumber\\
  &=& \frac{1}{2\sigma}\log\left(n^{2\sigma} \frac{(\bar p_\mathrm{ext}+\sigma)^{\bar p_\mathrm{ext}+\sigma}}{(\bar p_\mathrm{ext}-\sigma)^{\bar p_\mathrm{ext}-\sigma}}\right)-1,
\label{eq:g_1_uniform}
\end{eqnarray}
valid for $\sigma < p_\mathrm{ext}$, while for the case in which $\sigma=p_\mathrm{ext}$ the integral gives:
\begin{equation}
 G = \log(2n\bar p_\mathrm{ext})-1.
\end{equation}
Because the critical regime separates positive from negative population growth rates, it can be determined by solving the integral for $G=0$. The resulting equations describing the critical regime as a function of $\sigma$ and $\bar p_\mathrm{ext}$ can then be solved numerically. These solutions are plotted as a dashed line in Fig. \ref{fig:phases_pure}). 
In the case where $\sigma=p_\mathrm{ext}$, the critical point is located at $\bar p_\mathrm{ext}^c = e/2n$ (in our simulations we take $n=5$, and therefore $\bar p_\mathrm{ext}^c \simeq 0.27$).

\subsection*{Non-dispersal syndrome ($\alpha=0$)} As, in this case, all seeds are inbred, their quality parameter $q$ is reduced by a factor $(1-\delta)$ in each generation; i.e.  
\begin{equation} q(t) = (1-\delta) q(t-1) = ... = (1-\delta)^t q(0).  \end{equation} As the establishment probability is $ p_\mathrm{int} \times q(t)$, a dramatic (exponential) reduction of this probability can be expected in time for any inbreeding penalty factor $\delta>0$. In particular, assuming that all individuals in the community start with a common quality $q(0)$ and taking into account that $1+2+...+t-1=t(t-1)/2$ the population size at generation $t$ is: 
\begin{eqnarray}
 N(t) &=& n q(t-1) p_\mathrm{int} N({t-1}) \nonumber\\
 &=&  \left( \prod_{i=1}^t n (1-\delta)^{i-1} q(0) p_\mathrm{int} \right) N(0) \nonumber\\
 &=& (1-\delta)^{t(t-1)/2}\left( n q(0) p_\mathrm{int} \right)^t N(0).
\end{eqnarray}
From this equation, $G$ can be expressed as
\begin{equation}
G = \lim_{t\rightarrow\infty} \log\left((1-\delta)^{(t-1)/2} (n q(0) p_\mathrm{int} \right).
\label{eq:g_0}
\end{equation}
Note that, for any $\delta>0$, we always find that $G<0$ as $\lim_{t \rightarrow \infty} G= -\infty$.  Instead, in the absence of inbreeding depression ($\delta=0$), the critical transition point, $G=0$, is found at $p_\mathrm{int}^c=(nq(0))^{-1}$ (see the inset of Fig. \ref{fig:phases_pure}).

\subsection*{Mixed dispersal syndrome ($0<\alpha<1$)} The calculation for mixed dispersal strategies is slightly more complex than the two previous single-phenotype cases. For mixed dispersal, the quality parameter becomes a stochastic variable. The quality of non-dispersing seeds can be multiplicatively reduced sequentially (as for the $\alpha=0$ case above); however, in the case of mixed strategies, the quality parameter is reset to one whenever seeds are dispersed.  Despite the complexity of these dynamics, the distribution of quality parameters in the community, $\mathcal{P}(q,t)$, reaches a stationary state after a sufficiently large number of generations.  A practical way to estimate the value of the stationary quality parameter consists in computing its mean value, $\bar q$, over individuals and generations. To do that we fix it self-consistently by imposing its mean value to remain unaltered from one generation to the next. 

Defining $N^{\alpha=0}(t)$ and $N^{\alpha=1}(t)$ as the number of individuals grown from non-dispersed and dispersed seeds at generation $t$, respectively, the stationary quality can be expressed mathematically as 
\begin{eqnarray}
\bar q &=& \left\langle 1\cdot \frac{N^{\alpha=1}(t)}{N(t)} + (1-\delta)\bar q \frac{N^{\alpha=0}(t)}{N(t)} \right\rangle_t \nonumber\\
&=& \left\langle \frac{ \alpha n p_\mathrm{ext}(t) +(1-\alpha) n p_\mathrm{int} (1-\delta) {\bar q}^2 }{\alpha n p_\mathrm{ext}(t) + (1-\alpha) n p_\mathrm{int} \bar q } \right\rangle_t, \end{eqnarray}

\begin{figure}[htb!]
\centering \includegraphics[width=0.95\columnwidth]{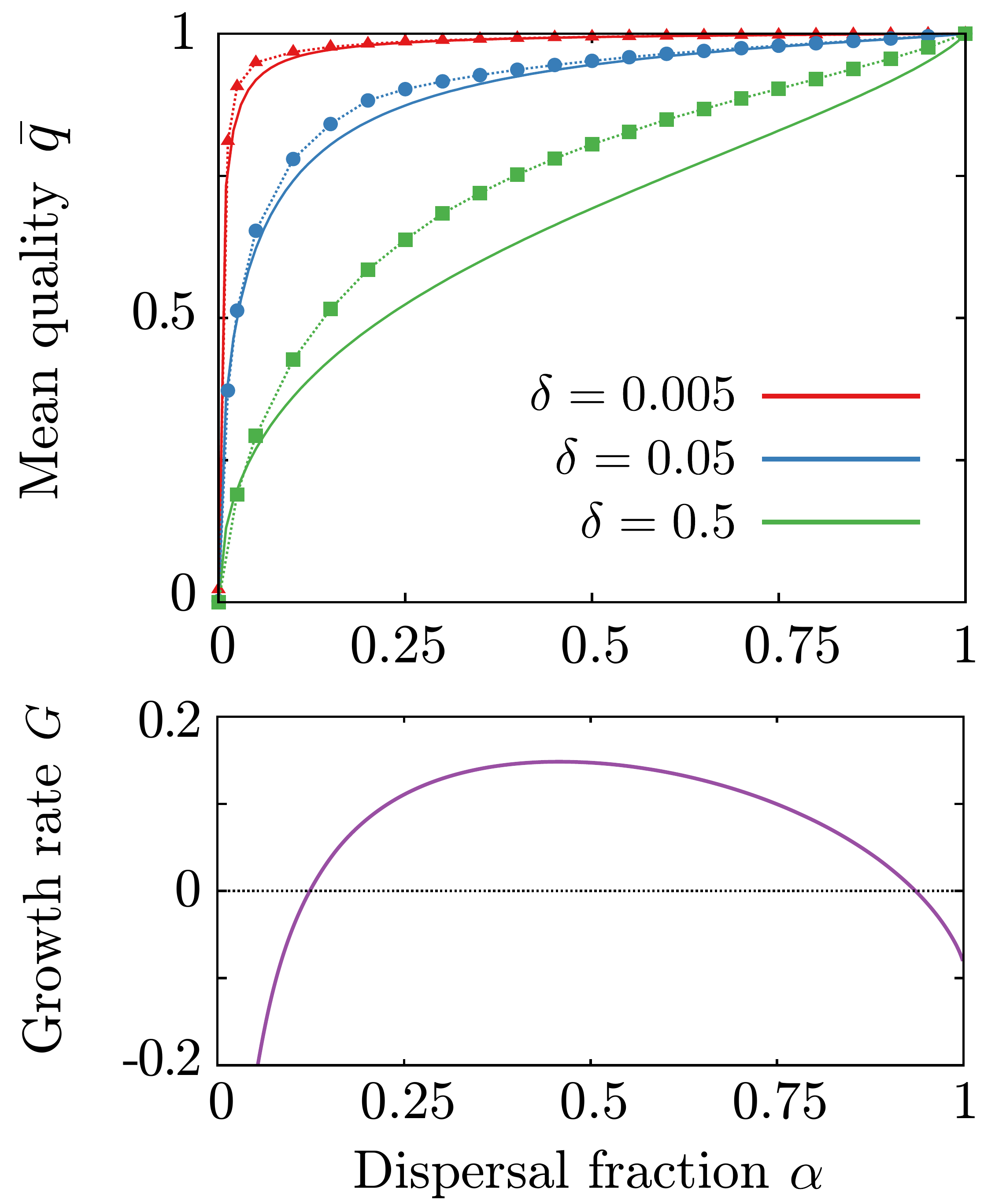}
\caption{\csentence{(Upper panel) Results of the self-consistent calculation used to estimate the averaged quality factor of seeds in a population with mixed dispersal strategies}.
After a sufficiently large number of generations, the quality parameter $q$ converges, on average, to a steady state value $\bar q$, which depends on $\alpha$ as shown for three different values of the inbreeding depression parameter $\delta$. The approximate values obtained within the self-consistent calculation (solid lines) are not far from the computational measured results (colored points), indicating that the approximations are sound. Observe that in the limit $\delta \rightarrow 0$ the curves converge to $\bar q = 1$ as expected. {\bf (Lower panel) Exponential growth rate $G$ as a function of the dispersal propensity parameter $\alpha$}. The growth rate, given by eq. \ref{eq:g_alpha_uniform}, interpolates non-linearly between $G(\alpha=0,\delta>0)=-\infty$ (eq. \ref{eq:g_0}) and $G(\alpha=1)$ (eq. \ref{eq:g_1_uniform}). As a consequence of its parabolic-like shape, $G(\alpha)$ intersects at zero, and intermediate values of the dispersal propensity parameter, $\alpha$ can be associated with $G(\alpha)>0$, even if $G(\alpha=0,1)<0$ (extinction); i.e., mixed dispersal can allow large populations to survive indefinitely. Parameters are as in Fig. \ref{fig:ext-times}: $p_\mathrm{int} = \bar p_\mathrm{ext} = \sigma = 0.25$, $n=5$ and, in the lower panel, $\delta=0.05$. Each point was computed in a community of $L=100$, averaging over the last $10^5/2$ generations of 10 independent simulations iterated for $10^5$ steps.}
\label{fig:theory}
\end{figure}

This equation can be solved --even if implicitly-- for $\bar q$ as a function of the dynamical parameters.  In the case where $p_\mathrm{ext}(t)$ is uniformly distributed in the range $[\bar p_\mathrm{ext}-\sigma, \bar p_\mathrm{ext}+\sigma]$, and assuming that time $t$ is large enough as to homogeneously sample all values of the $p_{\mathrm{ext}}$ distribution, the average quality can be calculated as:
\begin{eqnarray}
\bar q &=& 1 + \frac{1}{2\sigma\alpha} p_\mathrm{int}\left(1-(1-\delta)\bar q\right) \bar q (1-\alpha) \times \nonumber\\
& & \log\frac{p_\mathrm{int} \bar q (1-\alpha) + \alpha(\bar p_\mathrm{ext}-\sigma)}{p_\mathrm{int} \bar q (1-\alpha) + \alpha(\bar p_\mathrm{ext}+\sigma)}.
\label{eq:q_uniform}
\end{eqnarray}
The numerical solution of eq. \ref{eq:q_uniform} is represented in Fig. \ref{fig:theory} (upper panel) as a function of the parameter $\alpha$ for different values of inbreeding depression $\delta$. The numerical solutions are presented together with the comparable computational approximations. This figure shows that, as expected after multiple generations $\bar q(\alpha=0)=0$ and $\bar q(\alpha=1)=1$ corresponding to the quality of populations with single non-dispersal and dispersal syndromes respectively. We can see that, although we have used a simplistic approximation, the computed $\bar q$ constitutes a good estimate of the actual value determined from computer simulations of the full model.

Finally, using the inferred value of $\bar q=\bar q(\alpha,\bar p_\mathrm{ext},\sigma, p_\mathrm{int},\delta)$, we are able to compute $N(t)$:
\begin{eqnarray}
N(t) &=& \left(\alpha n p_\mathrm{ext}(t) + (1-\alpha)n \bar q p_\mathrm{int} \right) N({t-1}) = ... \nonumber\\
&=& \prod_{i=1}^t  \left(\alpha n p_\mathrm{ext}(i) + (1-\alpha)n \bar q p_\mathrm{int}  \right) N(0),
\end{eqnarray}
and from this
\begin{equation}
G = \left\langle \log  \left( \alpha n p_\mathrm{ext}(t) + (1-\alpha)n \bar q p_\mathrm{int}  \right) \right\rangle_t,  
\end{equation}
which in the case of a uniformly distributed environment in $[\bar p_\mathrm{ext}-\sigma, \bar p_\mathrm{ext}+\sigma]$ becomes
\begin{eqnarray}
  &&  G = -1+\frac{1}{2\sigma\alpha}\log\Bigg( n^{2\sigma\alpha} \times \\
  && \frac{\left[(1-\alpha) \bar q p_\mathrm{ext}+\alpha(\bar
      p_\mathrm{ext} + \sigma)\right]^{(1-\alpha) \bar q
      p_\mathrm{ext}+\alpha(\bar p_\mathrm{ext} +
      \sigma)}}{\left[(1-\alpha) \bar q p_\mathrm{ext}+\alpha(\bar
      p_\mathrm{ext} - \sigma)\right]^{(1-\alpha) \bar q
          p_\mathrm{ext}+\alpha(\bar p_\mathrm{ext} - \sigma)}}\Bigg)   \nonumber
\label{eq:g_alpha_uniform}
\end{eqnarray}
where $\bar q=\bar q(\alpha,\bar p_\mathrm{ext},\sigma, p_\mathrm{int},\delta)$ is the solution of eq. \ref{eq:q_uniform}. In Fig. \ref{fig:theory} (lower panel) the growth rate is plotted as a function of the dispersal fraction $\alpha$, for the same choice of parameters in Fig. \ref{fig:ext-times}, i.e. when both dispersing and non-dispersing syndromes are nonviable ($G(\alpha=0,1)<0$). As a consequence of its parabolic-like shape, $G$ intersects zero and becomes positive for intermediate values of the dispersal fraction. Therefore, the analytical prediction confirms that even when populations exhibiting either of the single phenotypes are bound to collapse, mixed dispersal syndromes can allow for long term population stability.


\begin{backmatter}


\section*{Author's contributions}
J.H., R.R.C. and M.A.M designed research; J.H. performed the simulations, J.H and M.A.M did the analytical calculations; J.H., R.R.C. and M.A.M. wrote the paper.

\section*{Acknowledgements}
We  are  grateful  to  S. Pigolotti  for  helpful  discussions  and  valuable  suggestions. 
M.A.M.  and  J.H.  acknowledge support  from  the  Spanish  MINECO  Excellence  project
FIS2013-43201-P.  J.H. thanks  the  support  of  the  University  of  Padova (PRAT2014-CPDA148037).
R.R.C. was funded by the European Commission (MC-IIF-2011-300026 ‘TEE-OFF’) and the Talentia program (Junta de Andalucia/ EC – FP7; grant ``Bet-hedging, trade-offs and the evolution of seed dispersal and dormancy'').

\bibliographystyle{bmc-mathphys} 

\newcommand{\BMCxmlcomment}[1]{}

\BMCxmlcomment{

<refgrp>

<bibl id="B1">
  <title><p>Maintenance of genetic heterogeneity</p></title>
  <aug>
    <au><snm>Dempster</snm><fnm>ER</fnm></au>
  </aug>
  <source>Cold Spring Harbor Symposia on Quantitative Biology</source>
  <pubdate>1955</pubdate>
  <volume>20</volume>
  <fpage>25</fpage>
  <lpage>-32</lpage>
</bibl>

<bibl id="B2">
  <title><p>Stochastic methods</p></title>
  <aug>
    <au><snm>Gardiner</snm><fnm>CW</fnm></au>
  </aug>
  <publisher>Berlin--Heidelberg--New York--Tokyo: Springer-Verlag</publisher>
  <pubdate>1985</pubdate>
</bibl>

<bibl id="B3">
  <title><p>Bet-hedging—a triple trade-off between means, variances and
  correlations</p></title>
  <aug>
    <au><snm>Starrfelt</snm><fnm>J</fnm></au>
    <au><snm>Kokko</snm><fnm>H</fnm></au>
  </aug>
  <source>Biological Reviews</source>
  <publisher>Wiley Online Library</publisher>
  <pubdate>2012</pubdate>
  <volume>87</volume>
  <issue>3</issue>
  <fpage>742</fpage>
  <lpage>-755</lpage>
</bibl>

<bibl id="B4">
  <title><p>Environmental and genetic sources of diversification in the timing
  of seed germination: implications for the evolution of bet
  hedging</p></title>
  <aug>
    <au><snm>Simons</snm><fnm>AM</fnm></au>
    <au><snm>Johnston</snm><fnm>MO</fnm></au>
  </aug>
  <source>Evolution</source>
  <publisher>Wiley Online Library</publisher>
  <pubdate>2006</pubdate>
  <volume>60</volume>
  <issue>11</issue>
  <fpage>2280</fpage>
  <lpage>-2292</lpage>
</bibl>

<bibl id="B5">
  <title><p>Evolutionary bet-hedging in the real world: empirical evidence and
  challenges revealed by plants</p></title>
  <aug>
    <au><snm>Childs</snm><fnm>DZ</fnm></au>
    <au><snm>Metcalf</snm><fnm>CJE</fnm></au>
    <au><snm>Rees</snm><fnm>M</fnm></au>
  </aug>
  <source>Proceedings of the Royal Society B: Biological Sciences</source>
  <publisher>The Royal Society</publisher>
  <pubdate>2010</pubdate>
  <fpage>rspb20100707</fpage>
</bibl>

<bibl id="B6">
  <title><p>Variable timing of reproduction in unpredictable environments:
  Adaption of flood plain plants</p></title>
  <aug>
    <au><snm>Satake</snm><fnm>A</fnm></au>
    <au><snm>Sasaki</snm><fnm>A</fnm></au>
    <au><snm>Iwasa</snm><fnm>Y</fnm></au>
  </aug>
  <source>Theoretical Population Biology</source>
  <publisher>Elsevier</publisher>
  <pubdate>2001</pubdate>
  <volume>60</volume>
  <issue>1</issue>
  <fpage>1</fpage>
  <lpage>-15</lpage>
</bibl>

<bibl id="B7">
  <title><p>Can spatial variation alone lead to selection for
  dispersal?</p></title>
  <aug>
    <au><snm>Hastings</snm><fnm>A</fnm></au>
  </aug>
  <source>Theoretical Population Biology</source>
  <publisher>Elsevier</publisher>
  <pubdate>1983</pubdate>
  <volume>24</volume>
  <issue>3</issue>
  <fpage>244</fpage>
  <lpage>-251</lpage>
</bibl>

<bibl id="B8">
  <title><p>Dispersal: risk spreading versus local adaptation</p></title>
  <aug>
    <au><snm>Kisdi</snm><fnm>{\'E}</fnm></au>
  </aug>
  <source>The American Naturalist</source>
  <publisher>JSTOR</publisher>
  <pubdate>2002</pubdate>
  <volume>159</volume>
  <issue>6</issue>
  <fpage>579</fpage>
  <lpage>-596</lpage>
</bibl>

<bibl id="B9">
  <title><p>Dispersal in stable habitats</p></title>
  <aug>
    <au><snm>Hamilton</snm><fnm>WD</fnm></au>
    <au><snm>May</snm><fnm>RM</fnm></au>
  </aug>
  <source>Nature</source>
  <pubdate>1977</pubdate>
  <volume>269</volume>
  <issue>5629</issue>
  <fpage>578</fpage>
  <lpage>-581</lpage>
</bibl>

<bibl id="B10">
  <title><p>Evolutionarily stable dispersal strategies</p></title>
  <aug>
    <au><snm>Comins</snm><fnm>HN</fnm></au>
    <au><snm>Hamilton</snm><fnm>WD</fnm></au>
    <au><snm>May</snm><fnm>RM</fnm></au>
  </aug>
  <source>Journal of Theoretical Biology</source>
  <publisher>Elsevier</publisher>
  <pubdate>1980</pubdate>
  <volume>82</volume>
  <issue>2</issue>
  <fpage>205</fpage>
  <lpage>-230</lpage>
</bibl>

<bibl id="B11">
  <title><p>The evolutionary ecology of seed heteromorphism</p></title>
  <aug>
    <au><snm>Venable</snm><fnm>DL</fnm></au>
  </aug>
  <source>American Naturalist</source>
  <publisher>JSTOR</publisher>
  <pubdate>1985</pubdate>
  <fpage>577</fpage>
  <lpage>-595</lpage>
</bibl>

<bibl id="B12">
  <title><p>Physiology and ecology of dispersal polymorphism in
  insects</p></title>
  <aug>
    <au><snm>Zera</snm><fnm>AJ</fnm></au>
    <au><snm>Denno</snm><fnm>RF</fnm></au>
  </aug>
  <source>Annual review of entomology</source>
  <publisher>Annual Reviews 4139 El Camino Way, PO Box 10139, Palo Alto, CA
  94303-0139, USA</publisher>
  <pubdate>1997</pubdate>
  <volume>42</volume>
  <issue>1</issue>
  <fpage>207</fpage>
  <lpage>-230</lpage>
</bibl>

<bibl id="B13">
  <title><p>Inbreeding depression and the evolution of dispersal rates: a
  multilocus model</p></title>
  <aug>
    <au><snm>Roze</snm><fnm>D</fnm></au>
    <au><snm>Rousset</snm><fnm>F</fnm></au>
  </aug>
  <source>The American Naturalist</source>
  <publisher>JSTOR</publisher>
  <pubdate>2005</pubdate>
  <volume>166</volume>
  <issue>6</issue>
  <fpage>708</fpage>
  <lpage>-721</lpage>
</bibl>

<bibl id="B14">
  <title><p>Joint evolution of dispersal and inbreeding load</p></title>
  <aug>
    <au><snm>Guillaume</snm><fnm>F</fnm></au>
    <au><snm>Perrin</snm><fnm>N</fnm></au>
  </aug>
  <source>Genetics</source>
  <publisher>Genetics Soc America</publisher>
  <pubdate>2006</pubdate>
  <volume>173</volume>
  <issue>1</issue>
  <fpage>497</fpage>
  <lpage>-509</lpage>
</bibl>

<bibl id="B15">
  <title><p>Avoiding inbreeding: at what cost?</p></title>
  <aug>
    <au><snm>Bengtsson</snm><fnm>BO</fnm></au>
  </aug>
  <source>Journal of Theoretical Biology</source>
  <publisher>Elsevier</publisher>
  <pubdate>1978</pubdate>
  <volume>73</volume>
  <issue>3</issue>
  <fpage>439</fpage>
  <lpage>-444</lpage>
</bibl>

<bibl id="B16">
  <title><p>The ecology and evolution of seed dispersal: a theoretical
  perspective</p></title>
  <aug>
    <au><snm>Levin</snm><fnm>SA</fnm></au>
    <au><snm>Muller Landau</snm><fnm>HC</fnm></au>
    <au><snm>Nathan</snm><fnm>R</fnm></au>
    <au><snm>Chave</snm><fnm>J</fnm></au>
  </aug>
  <source>Annual Review of Ecology, Evolution, and Systematics</source>
  <publisher>JSTOR</publisher>
  <pubdate>2003</pubdate>
  <fpage>575</fpage>
  <lpage>-604</lpage>
</bibl>

<bibl id="B17">
  <title><p>Spatial patterns of seed dispersal, their determinants and
  consequences for recruitment</p></title>
  <aug>
    <au><snm>Nathan</snm><fnm>R</fnm></au>
    <au><snm>Muller Landau</snm><fnm>HC</fnm></au>
  </aug>
  <source>Trends in ecology \& evolution</source>
  <publisher>Elsevier</publisher>
  <pubdate>2000</pubdate>
  <volume>15</volume>
  <issue>7</issue>
  <fpage>278</fpage>
  <lpage>-285</lpage>
</bibl>

<bibl id="B18">
  <title><p>How does it feel to be like a rolling stone? Ten questions about
  dispersal evolution</p></title>
  <aug>
    <au><snm>Ronce</snm><fnm>O</fnm></au>
  </aug>
  <source>Annual Review of Ecology, Evolution, and Systematics</source>
  <publisher>JSTOR</publisher>
  <pubdate>2007</pubdate>
  <fpage>231</fpage>
  <lpage>-253</lpage>
</bibl>

<bibl id="B19">
  <title><p>The ecology of amphicarpic plants</p></title>
  <aug>
    <au><snm>Cheplick</snm><fnm>GP</fnm></au>
  </aug>
  <source>Trends in ecology \& evolution</source>
  <publisher>Elsevier</publisher>
  <pubdate>1987</pubdate>
  <volume>2</volume>
  <issue>4</issue>
  <fpage>97</fpage>
  <lpage>-101</lpage>
</bibl>

<bibl id="B20">
  <title><p>Seed heteromorphism and the life cycle of plants: A literature
  review</p></title>
  <aug>
    <au><snm>Mand{\'a}k</snm><fnm>B</fnm></au>
  </aug>
  <source>Preslia</source>
  <publisher>The Czech Botanical Society</publisher>
  <pubdate>1997</pubdate>
  <volume>69</volume>
  <issue>2</issue>
  <fpage>129</fpage>
  <lpage>-159</lpage>
</bibl>

<bibl id="B21">
  <title><p>Ecological consequences and ontogeny of seed
  heteromorphism</p></title>
  <aug>
    <au><snm>Imbert</snm><fnm>E</fnm></au>
  </aug>
  <source>Perspectives in Plant Ecology, Evolution and Systematics</source>
  <publisher>Elsevier</publisher>
  <pubdate>2002</pubdate>
  <volume>5</volume>
  <issue>1</issue>
  <fpage>13</fpage>
  <lpage>-36</lpage>
</bibl>

<bibl id="B22">
  <title><p>The selection of cleistogamy and heteromorphic
  diaspores</p></title>
  <aug>
    <au><snm>Schoen</snm><fnm>DJ</fnm></au>
    <au><snm>Lloyd</snm><fnm>DG</fnm></au>
  </aug>
  <source>Biological Journal of the Linnean Society</source>
  <pubdate>1984</pubdate>
  <volume>23</volume>
  <issue>4</issue>
  <fpage>303</fpage>
  <lpage>-322</lpage>
</bibl>

<bibl id="B23">
  <title><p>A review and survey of basicarpy, geocarpy, and amphicarpy in the
  African and Madagascan flora</p></title>
  <aug>
    <au><snm>Barker</snm><fnm>NP</fnm></au>
  </aug>
  <source>Annals of the Missouri Botanical Garden</source>
  <publisher>JSTOR</publisher>
  <pubdate>2005</pubdate>
  <fpage>445</fpage>
  <lpage>-462</lpage>
</bibl>

<bibl id="B24">
  <title><p>Multiplicity in unity: plant subindividual variation and
  interactions with animals</p></title>
  <aug>
    <au><snm>Herrera</snm><fnm>CM</fnm></au>
  </aug>
  <publisher>Chicago--London: University of Chicago Press</publisher>
  <pubdate>2009</pubdate>
</bibl>

<bibl id="B25">
  <title><p>Pollination fluctuations drive evolutionary syndromes linking
  dispersal and mating system</p></title>
  <aug>
    <au><snm>Cheptou</snm><fnm>PO</fnm></au>
    <au><snm>Massol</snm><fnm>F</fnm></au>
  </aug>
  <source>The American Naturalist</source>
  <publisher>JSTOR</publisher>
  <pubdate>2009</pubdate>
  <volume>174</volume>
  <issue>1</issue>
  <fpage>46</fpage>
  <lpage>-55</lpage>
</bibl>

<bibl id="B26">
  <title><p>WHEN SHOULD WE EXPECT THE EVOLUTIONARY ASSOCIATION OF
  SELF-FERTILIZATION AND DISPERSAL?</p></title>
  <aug>
    <au><snm>Massol</snm><fnm>F</fnm></au>
    <au><snm>Cheptou</snm><fnm>PO</fnm></au>
  </aug>
  <source>Evolution</source>
  <publisher>Wiley Online Library</publisher>
  <pubdate>2011</pubdate>
  <volume>65</volume>
  <issue>5</issue>
  <fpage>1217</fpage>
  <lpage>-1220</lpage>
</bibl>

<bibl id="B27">
  <title><p>Clarifying Baker's law</p></title>
  <aug>
    <au><snm>Cheptou</snm><fnm>P O</fnm></au>
  </aug>
  <source>Annals of botany</source>
  <publisher>Annals Botany Co</publisher>
  <pubdate>2012</pubdate>
  <volume>109</volume>
  <issue>3</issue>
  <fpage>633</fpage>
  <lpage>-641</lpage>
</bibl>

<bibl id="B28">
  <title><p>Evolution of dispersal in spatially and temporally variable
  environments: The importance of life cycles</p></title>
  <aug>
    <au><snm>Massol</snm><fnm>F</fnm></au>
    <au><snm>D{\'e}barre</snm><fnm>F</fnm></au>
  </aug>
  <source>Evolution</source>
  <publisher>Wiley Online Library</publisher>
  <pubdate>2015</pubdate>
  <volume>69</volume>
  <issue>7</issue>
  <fpage>1925</fpage>
  <lpage>-1937</lpage>
</bibl>

<bibl id="B29">
  <title><p>The genetics of inbreeding depression</p></title>
  <aug>
    <au><snm>Charlesworth</snm><fnm>D</fnm></au>
    <au><snm>Willis</snm><fnm>JH</fnm></au>
  </aug>
  <source>Nature Reviews Genetics</source>
  <publisher>Nature Publishing Group</publisher>
  <pubdate>2009</pubdate>
  <volume>10</volume>
  <issue>11</issue>
  <fpage>783</fpage>
  <lpage>-796</lpage>
</bibl>

<bibl id="B30">
  <title><p>How to simulate the quasistationary state</p></title>
  <aug>
    <au><snm>Oliveira</snm><fnm>MM</fnm></au>
    <au><snm>Dickman</snm><fnm>R</fnm></au>
  </aug>
  <source>Physical Review E</source>
  <publisher>APS</publisher>
  <pubdate>2005</pubdate>
  <volume>71</volume>
  <issue>1</issue>
  <fpage>016129</fpage>
</bibl>

<bibl id="B31">
  <title><p>Frequency dependence versus optimization</p></title>
  <aug>
    <au><snm>Kisdi</snm><fnm>{\'E}</fnm></au>
  </aug>
  <source>Trends in ecology \& evolution</source>
  <publisher>Elsevier Current Trends</publisher>
  <pubdate>1998</pubdate>
  <volume>13</volume>
  <issue>12</issue>
  <fpage>508</fpage>
</bibl>

<bibl id="B32">
  <title><p>Genetic Algorithms in Search, Optimization, and Machine
  Learning</p></title>
  <aug>
    <au><snm>Goldberg</snm><fnm>DE</fnm></au>
  </aug>
  <publisher>Boston: Addison-Wesley Professional</publisher>
  <pubdate>1989</pubdate>
</bibl>

<bibl id="B33">
  <title><p>Complex and adaptive dynamical systems: A primer</p></title>
  <aug>
    <au><snm>Gros</snm><fnm>C</fnm></au>
  </aug>
  <publisher>Berlin--Heidelberg: Springer</publisher>
  <pubdate>2013</pubdate>
</bibl>

<bibl id="B34">
  <title><p>Populations can persist in an environment consisting of sink
  habitats only</p></title>
  <aug>
    <au><snm>Jansen</snm><fnm>VA</fnm></au>
    <au><snm>Yoshimura</snm><fnm>J</fnm></au>
  </aug>
  <source>Proceedings of the National Academy of Sciences</source>
  <publisher>National Acad Sciences</publisher>
  <pubdate>1998</pubdate>
  <volume>95</volume>
  <issue>7</issue>
  <fpage>3696</fpage>
  <lpage>-3698</lpage>
</bibl>

<bibl id="B35">
  <title><p>Metapopulation dynamics: brief history and conceptual
  domain</p></title>
  <aug>
    <au><snm>Hanski</snm><fnm>I</fnm></au>
    <au><snm>Gilpin</snm><fnm>M</fnm></au>
  </aug>
  <source>Biological journal of the Linnean Society</source>
  <publisher>Wiley Online Library</publisher>
  <pubdate>1991</pubdate>
  <volume>42</volume>
  <issue>1-2</issue>
  <fpage>3</fpage>
  <lpage>-16</lpage>
</bibl>

<bibl id="B36">
  <title><p>Metapopulation ecology</p></title>
  <aug>
    <au><snm>Hanski</snm><fnm>I</fnm></au>
  </aug>
  <publisher>Oxford--New~York: Oxford University Press Oxford</publisher>
  <pubdate>1999</pubdate>
  <volume>232</volume>
</bibl>

<bibl id="B37">
  <title><p>Multiple risk reduction mechanisms: can dormancy substitute for
  dispersal?</p></title>
  <aug>
    <au><snm>Snyder</snm><fnm>RE</fnm></au>
  </aug>
  <source>Ecology Letters</source>
  <publisher>Wiley Online Library</publisher>
  <pubdate>2006</pubdate>
  <volume>9</volume>
  <issue>10</issue>
  <fpage>1106</fpage>
  <lpage>-1114</lpage>
</bibl>

<bibl id="B38">
  <title><p>Leaving home aint easy non-local seed dispersal is only
  evolutionarily stable in highly unpredictable environments</p></title>
  <aug>
    <au><snm>Snyder</snm><fnm>RE</fnm></au>
  </aug>
  <source>Proceedings of the Royal Society B: Biological Sciences</source>
  <publisher>The Royal Society</publisher>
  <pubdate>2011</pubdate>
  <volume>278</volume>
  <issue>1706</issue>
  <fpage>739</fpage>
  <lpage>-744</lpage>
</bibl>

<bibl id="B39">
  <title><p>Evolutionary stability of plant communities and the maintenance of
  multiple dispersal types</p></title>
  <aug>
    <au><snm>Ludwig</snm><fnm>D</fnm></au>
    <au><snm>Levin</snm><fnm>SA</fnm></au>
  </aug>
  <source>Theoretical Population Biology</source>
  <publisher>Elsevier</publisher>
  <pubdate>1991</pubdate>
  <volume>40</volume>
  <issue>3</issue>
  <fpage>285</fpage>
  <lpage>-307</lpage>
</bibl>

<bibl id="B40">
  <title><p>Evolution of the Magnitude and Timing of Inbreeding Depression in
  Plants</p></title>
  <aug>
    <au><snm>Husband</snm><fnm>BC</fnm></au>
    <au><snm>Schemske</snm><fnm>DW</fnm></au>
  </aug>
  <source>Evolution</source>
  <pubdate>1996</pubdate>
  <volume>50</volume>
  <issue>1</issue>
  <fpage>54</fpage>
  <lpage>-70</lpage>
</bibl>

<bibl id="B41">
  <title><p>Dispersal in patchy environments: the effects of temporal and
  spatial structure</p></title>
  <aug>
    <au><snm>Cohen</snm><fnm>D</fnm></au>
    <au><snm>Levin</snm><fnm>SA</fnm></au>
  </aug>
  <source>Theoretical Population Biology</source>
  <publisher>Elsevier</publisher>
  <pubdate>1991</pubdate>
  <volume>39</volume>
  <issue>1</issue>
  <fpage>63</fpage>
  <lpage>-99</lpage>
</bibl>

<bibl id="B42">
  <title><p>Nonequilibrium phase transitions in lattice models</p></title>
  <aug>
    <au><snm>Marro</snm><fnm>J</fnm></au>
    <au><snm>Dickman</snm><fnm>R</fnm></au>
  </aug>
  <publisher>Cambridge: Cambridge University Press</publisher>
  <pubdate>2005</pubdate>
</bibl>

</refgrp>
} 

\end{backmatter}
\end{document}